\documentclass[usenatbib,fleqn]{mnras}
\usepackage{graphicx}
\usepackage[fleqn]{amsmath}
\usepackage{multicol}
\usepackage[english]{babel}
\usepackage{epstopdf}
\usepackage{calrsfs}
\DeclareMathAlphabet{\pazocal}{OMS}{zplm}{m}{n}
\newcommand{\Pa}{\mathcal{P}}

\newcommand{\mpchi}{\,h^{-1}{\rm {Mpc}}}

\newcommand{\msun}{\,h^{-1}{\rm M_{\sun}}}
\newcommand{\Mh}{M_{\rm h}}
\newcommand{\rp}{r_{\rm p}}
\newcommand{\ngg}{n_{\rm g}}

\newcommand{\Vmax}{$V_{\rm max}$}
\newcommand{\Vpeak}{$V_{\rm peak}$}
\newcommand{\Vacc}{$V_{\rm acc}$}
\newcommand{\Macc}{$M_{\rm acc}$}
\newcommand{\fsig}{$f\sigma_8$}
\newcommand{\hatfs}{$\widehat{f\sigma_8}$}

\usepackage{footmisc}

\title[Effects of Assembly Bias on Growth Rate from RSD]{The Effects of Galaxy Assembly Bias on the Inference of Growth Rate from Redshift-Space Distortions}

\author[K.S. McCarthy, Z. Zheng, and H. Guo]{
\parbox{\textwidth}{
Kevin S. McCarthy$^{1}$\thanks{E-mail: kevin.mccarthy@utah.edu}, 
Zheng Zheng$^{1}$\thanks{E-mail: zhengzheng@astro.utah.edu},
and 
Hong Guo$^{2}$}\\
\\
$^{1}$ Department of Physics and Astronomy, University of Utah, UT 84112, USA\\
$^{2}$ Key Laboratory for Research in Galaxies and Cosmology, Shanghai Astronomical Observatory, Shanghai 200030, China
}

\begin{document}
\label{firstpage} \pagerange{\pageref{firstpage}--\pageref{lastpage}}

\maketitle

\begin{abstract}
The large-scale redshift-space distortion (RSD) in galaxy clustering can probe \fsig, a combination of cosmic structure growth rate and matter fluctuation amplitude, which can constrain dark energy models and test theories of gravity. While the RSD on small scales (e.g. a few to tens of $\mpchi$) can further tighten the \fsig\ constraints,  galaxy assembly  bias,  if  not correctly  modeled,  may introduce systematic uncertainties. Using a mock galaxy catalogue with built-in assembly bias, we perform a preliminary study on how assembly bias may affect the \fsig\ inference. We find good agreement on scales down to 8--9$\mpchi$ between a \fsig\ metric from the redshift-space two-point correlation function with the central-only mock catalogue and that with the shuffled catalogue free of assembly bias, implying that \fsig\ information can be extracted on such scales even with assembly bias. We then apply the halo occupation distribution (HOD) and three subhalo clustering and abundance matching (SCAM) models to model the redshift-space clustering with the mock. Only the SCAM model based on \Vpeak\ (used to create the mock) can reproduce the \fsig\ metric, and the other three could not. However, the \fsig\ metrics determined from central galaxies from all the models are able to match the expected one down to 8$\mpchi$. Our results suggest that halo models with no or an incorrect assembly bias prescription could still be used to model the RSD down to scales of $\sim 8\mpchi$ to tighten the \fsig\ constraint, with a sample of central galaxies or with a flexible satellite occupation prescription.

\end{abstract}
\begin{keywords}
galaxies: distances and redshifts 
-- galaxies: haloes 
-- galaxies: statistics 
-- cosmology: large-scale structure of Universe 
-- cosmology: cosmological parameters
\end{keywords}

\section{Introduction}\label{sec:intro}

Constraining the nature of the accelerated expansion of the universe has become  the major driver of ongoing and 
forthcoming galaxy redshift surveys, such as the Sloan Digital Sky Survey III/IV (SDSS III/IV; 
\citealt{Dawson13,Abolfathi18}), 
Dark Energy Spectroscopic Instrument (DESI; \citealt{DESI16}), 
Wide Field Infrared Survey Telescope (WFIRST; \citealt{Spergel15}), 
and Euclid \citep{Laureijs11}. One promising probe is through the motion of
galaxies, which causes an additional redshift 
component besides the cosmological redshifts from the uniform expansion of the universe, leaving a pattern of redshift-space \citep{Sargent77} distortions (RSD) in galaxy clustering. Since the large-scale galaxy motion is sourced by density fluctuation, RSD can 
probe the cosmic structure growth rate, more accurately, a combination of the growth rate and the density fluctuation amplitude, denoted by $f\sigma_8$. Here $f\equiv\partial \ln\delta/\partial \ln a$ is the linear growth rate,
with $\delta$ the linear density fluctuation amplitude and $a$ the scale factor, and $\sigma_8$ is the r.m.s. 
matter density fluctuation on scales of 8$\mpchi$.  The growth rate is determined by the way gravity
works, i.e. closely related to the equation of state and evolution of dark energy and the theory of gravity.

Most of the constraints on the growth rate use large-scale (above tens of Mpc) redshift-space galaxy clustering 
\citep[e.g.][]{Blake13,Alam17,Ruggeri18}, 
as it is relatively easy to model the RSD effect in linear or mildly nonlinear regime. The uncertainties
are generally large, limiting the constraining power of RSD on models of dark energy and gravity. The intermediate-
and small-scale redshift-space clustering has high statistical power, which, if used, can in principle greatly 
tighten the constraints on the linear growth rate to a precision at the level of a few per cent \citep[e.g.][]{Reid14,Dawson16,Zhai18}. 
In this paper, we investigate a potential 
systematic effect on $f\sigma_8$ constraints when extending the RSD modelling into intermediate and small scales, namely
the galaxy assembly bias effect. 
In modelling galaxy clustering, the halo model \citep[e.g.][]{Cooray02} is widely used to connect galaxy distribution to 
the underlying dark matter distribution that encodes cosmological information. The commonly adopted frameworks of
the galaxy-halo connection include the halo occupation distribution (HOD; e.g. \citealt{Berlind03}) and the 
conditional luminosity function (CLF; e.g. \citealt{Yang03}). One implicit assumption in such frameworks is that 
the statistical properties of galaxies inside haloes only depend on halo mass and not on other halo properties, such
as those related to the assembly history or environment of haloes. However, it has been established that in addition
to halo mass the clustering of haloes has dependencies on halo assembly properties \citep[e.g.][]{Gao05,Gao07,Xu18,Mansfield19}, 
although the nature of such halo assembly bias is still under investigation \citep[e.g.][]{Dalal08,Ramakrishnan19}. 
If in haloes of fixed mass galaxy properties are affected by halo assembly, we would have galaxy assembly bias, which would invalidate the assumption of the mass-only dependence in the above models  of galaxy-halo connection. If galaxy assembly bias is strong and
not accounted for in modelling galaxy clustering, it would lead to systematic effects in galaxy-halo connection
and in cosmological constraints \citep[e.g.][]{Zentner14}. It is necessary to study how galaxy assembly bias 
affects the use of the RSD effect to constrain the linear growth rate.

Our investigation is motivated by the work of \citet{McEwen16}, who study the assembly bias effect on inferring the 
matter clustering from galaxy clustering and weak lensing observation, using a mock galaxy catalogue \citep{Hearin13} 
with significant galaxy assembly bias. They find that the standard HOD is able to sufficiently model galaxy-galaxy and 
galaxy-matter two-point correlation functions (2PCFs) into nonlinear regime, giving a good description of 
galaxy-matter cross-correlation coefficient and accurately recovering the matter clustering down to a scale of 
$\sim 2\mpchi$. In other words, an intrinsically incorrect model that fits the data well can lead to the correct 
inference of the matter correlation function down to small scales. 

In parallel to the \citet{McEwen16} study, we investigate how well the HOD and other halo models work in inferring the 
structure growth rate from the RSD effect and down to what scales the redshift-space clustering measurements can still be 
used to contribute to the constraints, with the existence of galaxy assembly bias effect. On large scales, the assembly bias is expected to have little effect on the inference of $f\sigma_8$. Conceptually, we can see this by considering the linear RSD regime and applying the Kaiser formula \citep{Kaiser87}. The shape of the galaxy redshift-space 2PCF or power spectrum provides the constraint on $f/b$, while the amplitude encodes the information of $b\sigma_8$, where $b$ is the linear galaxy bias. In combination, a constraint on $f\sigma_8$ is obtained. Assembly bias would change the value of $b$, thus each of the constraints on $f/b$ and $b\sigma_8$, but not their combination (product), which is independent of $b$. However, on intermediate and small scales, the situation is not clear, which is what we intend to investigate.

The paper is organised as follows. In section~\ref{sec:methods}, we present the mock galaxy catalogues and our 
general methods in the study. In section~\ref{sec:Results}, we provide the results from our analyses. Finally,
in section~\ref{sec:discussion}, we give a summary and discuss the implication for RSD measurements, as well as 
potential routes for progress.

\section{Methodology}\label{sec:methods}
\subsection{Mock galaxy catalogues}\label{sec:obs}

We base our study on a mock galaxy catalogue constructed by \citet{Hearin13} (hereafter HW13) through populating 
galaxies into the dark matter haloes and subhaloes  identified in the Bolshoi N-body 
simulation\footnote{https://www.cosmosim.org} 
\citep{Klypin11}. The simulation adopts a spatially-flat cosmological model with $\Omega_{\rm m}$=0.27, 
$\Omega_{\Lambda}$=0.73, $\Omega_{\rm b}$=0.0469, $n_{\rm s}=0.95$, $\sigma_8$=0.82, and $H_0=100h\,{\rm km\, s^{-1}Mpc^{-1}}$ with $h$=0.7. It simulates the evolution of the 
matter density field with 2048$^3$ dark matter particles in a cubic box of side length 250$\mpchi$, with particle mass
$1.35 \times 10^8 \msun$ and force resolution of 1.0$h^{-1}$kpc (physical). Dark matter haloes are identified with 
the {\sc rockstar} halo finder \citep{Behroozi13a} and merger-sensitive parameters found with the assistance of 
{\sc consistent trees} \citep{Behroozi13b}.

The HW13 mock galaxy catalogue is constructed with the subhalo abundance matching (SHAM) method, through performing 
abundance matching (with scatter) between galaxy 
luminosity and \Vpeak, where \Vpeak\ is the highest maximum circular velocity (\Vmax) of a dark matter halo/subhalo 
during its assembly history. The halo \Vpeak\ property has been shown to be strongly correlated with galaxy stellar mass \citep{Reddick13,Xu19}. Galaxies put to the centres (places of lowest potential) of host haloes are central 
galaxies, and those to subhaloes are satellites. The halo assembly bias effect in terms of \Vpeak\ is therefore 
inherited by galaxy luminosity. While we focus our analyses using galaxy samples defined by $r$-band luminosity, 
HW13 also assign colours ($g-r$) to galaxies with an age-matching technique. The projected 2PCFs of HW13 mock galaxies 
with various luminosity thresholds are found to reasonably reproduce those measured from the SDSS DR7 data. We study
three luminosity threshold samples with ${\rm M}_r-5\log h <$ -19, -20, and -21, which are named Mr19, Mr20, and 
Mr21, respectively. The numbers of total (central) galaxies in the three samples are  244,784 (181,593), 96,646 
(74,242) and 17,268 (14,062), with corresponding number densities of $n_g\sim 1.6\times 10^{-2}$, $6.2\times 10^{-3}$ and $1.1\times 10^{-3}h^3{\rm Mpc}^{-3}$.

For comparison purpose, we also construct a control mock catalogue to eliminate the large-scale galaxy assembly bias,
following \citet{Croton07}. In short, we divide haloes into narrow mass bins of width 0.07 dex, and in each bin we
shuffle the galaxy contents (central and satellite galaxies) among haloes, including those with no galaxies above the 
luminosity thresholds. The satellite phase-space coordinates with respect to the central galaxy are preserved for 
each system. We note that the shuffled mock reserves any assembly effect inside each individual halo and keeps the one-halo term of the 2PCF. However, the construction completely erases the assembly bias signal present in the two-halo central-central term that encodes the cosmological information, which serves our purpose in this paper (see also \citealt{Zentner14}).

\subsection{$f\sigma_8$ estimator and measurements}\label{sec:fsig8}

To constrain $f\sigma_8$ from the RSD effect, the ideal route is to perform fitting or Markov Chain Monte Carlo (MCMC)
analysis to the measurements of the redshift-space galaxy clustering on all scales with a model (e.g. HOD). The model
should account for changes in cosmology (and theory of gravity), which lead to changes in the halo population and 
halo motion. Such a nontrivial task 
could be done by resorting to efficient methods of galaxy clustering modelling \citep[e.g.][]{Zheng16} or emulators 
\citep[e.g.][]{Wibking17,Zhai18}. While this should be developed to model observational data, it is beyond the scope of our investigation presented here for the effect of assembly bias on the 
$f\sigma_8$ inference. We rely on a simulation with fixed cosmology, and therefore we construct a quantity that 
is characterised by and closely related to $f\sigma_8$ and see how assembly bias affects it, following the spirit in 
\citet{McEwen16}.

\citet{Percival09} propose an estimator of \fsig\ based on the redshift-space galaxy power spectrum in the Kaiser 
regime. As our analyses are based on the redshift-space 2PCFs, we follow the procedure in \citet{Percival09} to 
construct the estimator in terms of the multipoles of the redshift-space 2PCF. The three non-vanishing terms of the
linear redshift-space 2PCF multipoles read \citep{Hamilton92}
\begin{eqnarray}
\xi_0 & = & \left(1+\frac{2}{3}\beta+\frac{1}{5}\beta^2\right)\xi_{\rm gg}, \label{eqn:1}\\
\xi_2 & = & \left(\frac{4}{3}\beta+\frac{4}{7}\beta^2\right) \left(\xi_{\rm gg}-\bar{\xi}_{\rm gg}\right), \\
\xi_4 & = & \frac{8}{35}\beta^2 \left(\xi_{\rm gg}+\frac{5}{2}\bar{\xi}_{\rm gg}-\frac{7}{2}\bar{\bar{\xi}}_{\rm gg}\right),
\label{eqn:corrarray}
\end{eqnarray}
where $\xi_{\rm gg}=\xi_{\rm gg}(r)$ is the real-space galaxy 2PCF and $\bar{\xi}_{\rm gg}$ and $\bar{\bar{\xi}}_{\rm gg}$ are 
averages of $\xi_{\rm gg}$ weighted by $r^3$ and $r^5$ (see Appendix~\ref{sec:A}), respectively. 
If we denote the matter 2PCF as $\xi_{\rm mm}$, in the linear regime we have $\xi_{\rm gg}=b^2\xi_{\rm mm}$, where $b$ is 
the galaxy bias factor. We can write $\xi_{\rm mm}=\sigma_8^2\xi^{\prime}_{\rm mm}$. 
That is, $\xi^{\prime}_{\rm mm}$ represents the shape of the matter 2PCF
with $\sigma_8$ set to unity. We then reformulate the multipoles as
\begin{eqnarray}
\tilde{\xi}_0 & \equiv &\frac{\xi_0}{\xi^{\prime}_{\rm mm}}=\left(1+\frac{2}{3}\beta+\frac{1}{5}\beta^2\right)(b\sigma_8)^2, \label{eqn:modcorrarray1}\\
\tilde{\xi}_2 & \equiv & \frac{\xi_2}{\xi^{\prime}_{\rm mm}-\bar{\xi}^{\prime}_{\rm mm}}=\left(\frac{4}{3}\beta+\frac{4}{7}\beta^2\right)(b\sigma_8)^2, \label{eqn:modcorrarray2}\\
\tilde{\xi}_4 & \equiv & \frac{\xi_4}{\xi^{\prime}_{\rm mm}+(5/2)\bar{\xi}^{\prime}_{\rm mm}-(7/2)\bar{\bar{\xi}}^{\prime}_{\rm mm}}=\frac{8}{35}\beta^2(b\sigma_8)^2. \label{eqn:6}
\label{eqn:modcorrarray3}
\end{eqnarray}
Following the same reasoning as in \citet{Percival09}, we reach the following estimator of \fsig\ with 2PCF multipoles,
\begin{equation}
\widehat{f\sigma_8}^2=
\frac{7}{48}\left[5\left(7\tilde{\xi}_0+\tilde{\xi}_2\right)-\sqrt{35}\left(35{\tilde{\xi}_0}^2+10\tilde{\xi}_0\tilde{\xi}_2-7{\tilde{\xi}_2}^2\right)^{1/2}\right].
\label{eq:estimator}
\end{equation}
Compared to equation~(29) in \citet{Percival09}, the above equation gives the configuration space counterpart of their expression 
in Fourier space.

As discussed in \citet{Percival09}, the estimator shows dependence on scale and approaches to \fsig\ only on 
substantially large scales (e.g. $k<0.05 h{\rm Mpc}^{-1}$). The deviation reflects the fact that Kaiser formula
is not accurate towards smaller, quasi-linear scales, in particular, when the Finger-of-God (FOG) effect \citep{Jackson72,Tully78} starts to dominates.
The estimator can be improved by introducing a FOG model \citep{Percival09}. In configuration space, the FOG is confined to 
scales with small projected pair separations, and to mitigate the effect we can compute the multipoles by excluding the 2PCF 
measurements within projected separation of 2$\mpchi$ (see Appendix~\ref{sec:B} for more 
details). As shown later, the \hatfs\ estimator is not expected to be constant even in the weakly 
nonlinear regime, where Kaiser formula is not accurate \citep[e.g.][]{Reid11}. Nevertheless, with the FOG mitigated the \hatfs\
estimator characterises the (scale-dependent) shape and amplitude of the RSD, which is determined by the velocity 
field and thus encodes the information of the linear growth rate \fsig. For a given galaxy sample, 
the constraints on \fsig\ can be inferred through 
fitting the \hatfs\ curve. In such a sense, we  can extend the estimator \hatfs\ to small scales (a few 
Mpc) that probes the halo motion. Our investigation focus on studying how assembly bias affects the \hatfs\ curve and whether the
commonly used halo models are able to provide a good description of the curve for \fsig\ inference.

To compute the multipoles in equations~(\ref{eqn:1})--(\ref{eqn:6}) with a mock galaxy catalogue, 
we first measure the redshift-space 2PCF as a function of $s=|\mathbf{s}|$ and $\mu=\cos\theta$, with $\theta$ angle 
between the line connecting a galaxy pair and the line of sight. It is done with the \citet{Peebles74} estimator for
the periodic box, $\xi(s,\mu)=DD(s,\mu)/RR(s,\mu)-1$, where $DD$ and $RR$ are the counts of data-data and 
random-random pairs in each ($s,\mu$) bin.
The multipoles are then computed as 
\begin{equation}
\xi_l=(2l+1)\sum_i\xi(s,\mu_i) \Pa_l(\mu_i)\Delta\mu_i, \quad\quad l=0,2,4,
\label{eq:expand}
\end{equation}
with $\Pa_l$ being the $l$-th Legendre polynomial. We use 20 $\mu$ bins from 0 to 1 ($\Delta\mu=0.05$), and the 
pair separation is divided into 27 equal logarithmic bins ($\Delta \log s=0.1$ dex) from $\log s=-1$ to $\log s=1.7$ (with $s$
in units of $\mpchi$). We use the mass particles in the Bolshoi simulation to compute the matter 2PCF $\xi^\prime_{mm}$ and 
the volume average ($\bar{\xi}^\prime_{mm}$) in equations~(\ref{eqn:modcorrarray1}) and (\ref{eqn:modcorrarray2}), and the 
method to compute the latter can be found in Appendix~\ref{sec:A}. As mentioned above, to reduce the FOG effect, 
we also compute the multipoles from computing the modified multipoles by excluding pairs with transverse separations 
smaller than 2$\mpchi$, which is detailed in Appendix~\ref{sec:B}.

Finally, for modelling the clustering, we include the projected 2PCF $w_{\rm p}$. 
It is measured by calculating the 2PCF as a function of the transverse and line-of-sight separations $\rp$ 
and $r_\pi$, again using the \citet{Peebles74} estimator, $\xi(\rp,r_\pi)=DD(\rp,r_\pi)/RR(\rp,r_\pi)-1$. 
The transverse separation $\rp$ is logarithmically binned in the same way as $s$, from $\log \rp=-1$ to $\log \rp=1.7$ 
with $27$ equal logarithmic bins ($\Delta \log \rp=0.1$ dex). The line-of-sight separation $r_\pi$ is linearly binned 
over $0$--$40$$\mpchi$ with bin width of $\Delta r_\pi=2\mpchi$. The projected 2PCF $w_{\rm p}(\rp)$ is then the integration 
of $\xi(\rp,r_\pi)$ along the $r_\pi$ direction. 

For 2PCF measurement of either $\xi(s,\mu)$ or $\xi(\rp,r_\pi)$, we choose each of the three principal axes of the simulation 
box as the line-of-sight direction and take the average of the three measurements as the measurement to be used. 
\begin{figure*}
\includegraphics[width=\textwidth]{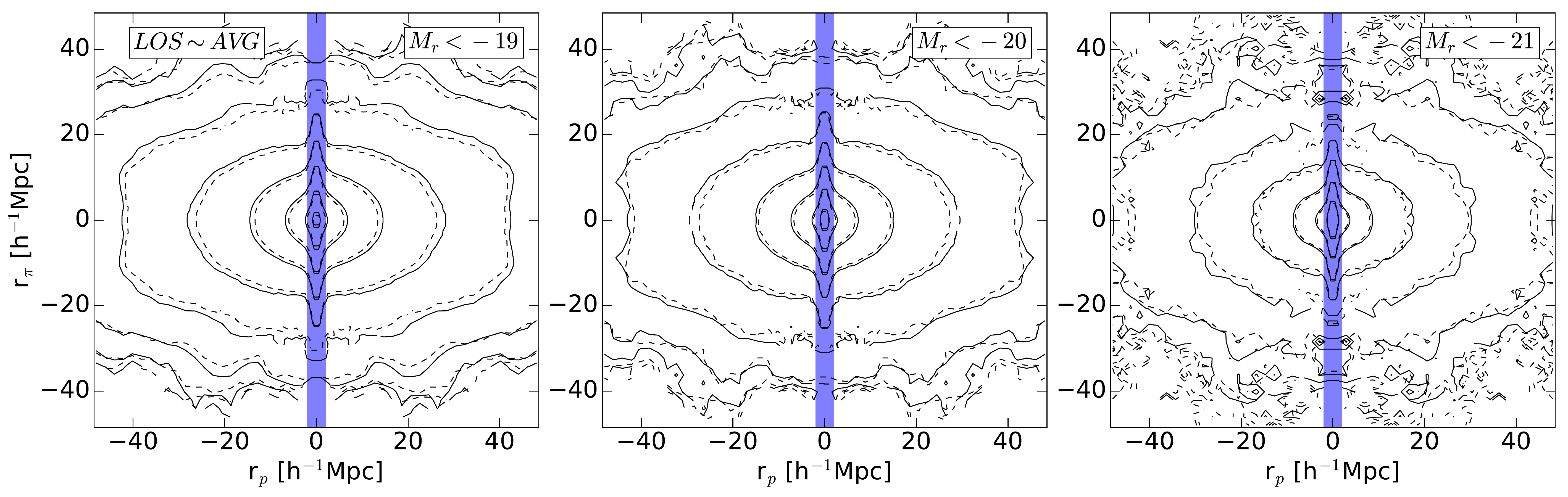}
\caption{
Redshift-space 2PCFs $\xi(\rp,r_\pi)$ for three luminosity-threshold mock galaxy samples. In each panel, the solid 
contours correspond to the measurements from the HW13 mock, while the dashed ones are from the shuffled mock that
removes the large-scale assembly bias. The vertical band indicates the region excluded 
to reduce the FOG effect on the calculation of the redshift-space 2PCF multipoles. See text for detail.
}
\label{fig:rppi}
\end{figure*}

\subsection{Halo modelling}\label{sec:modfits}

To study whether commonly used halo models are able to describe the intermediate- and small-scale RSD to help infer
unbiased \fsig\ for a galaxy population with assembly bias, we perform halo modelling of the RSD measurements from the
HW13 mock catalogue. As the purpose of the investigations is to see whether useful cosmological information can be extracted even with models with no or incomplete assembly bias effect, we are not particularly interested in studying the model parameters and discussing how well the galaxy-halo relation is reproduced, following the philosophy in \citet{McEwen16}. We consider two types of models, the HOD model and the sub-halo clustering abundance matching (SCAM; \citealt{Guo16}). The modelling results are used to produce the \hatfs\ curve to compare to the one measured from the HW13 mock.

For the HOD model, we adopt the standard five-parameter description \citep{Zheng05} for the mean occupation function of galaxies above a luminosity threshold.  The mean occupation function $\langle N_{\rm cen}(\Mh)\rangle$ for central galaxies is a step-like function with a characteristic mass $M_{\rm min}$ and a transition width $\sigma_{\log M}$, and that for satellites follows a power law with an index $\alpha$, amplitude parameter $M_1^\prime$,  and a low-mass cutoff determined by $M_0$, respectively,
\begin{eqnarray}
\langle N_{\rm cen}(\Mh) \rangle & = & \frac{1}{2}\left[1+{\rm erf}\left(\frac{\log M_h-\log M_{\rm min}}{\sigma_{\log M}}\right)\right], \label{eqn:hodcen}\\
\langle N_{\rm sat}(\Mh)\rangle  & = & \langle N_{\rm cen}(\Mh)\rangle \left(\frac{\Mh-M_0}{M_1^\prime}\right)^\alpha.
\label{eqn:hodsat}
\end{eqnarray}

Central galaxies are put at the halo centres and random particles in haloes are assigned as satellites \citep[see][]{Guo15a}. As the redshift-space clustering is to be modelled, we also include two additional parameters $\alpha_c$ and $\alpha_s$ to describe the central and satellite velocity bias \citep{Guo15a}. That is, the central galaxies can move with respect to the halo centres with a velocity dispersion of $\alpha_c$ times that of the dark matter inside haloes; the velocity dispersion of satellites is $\alpha_s$ times that of the dark matter inside haloes. 

Like the SHAM method \citep{Vale04,Conroy06}, the SCAM model makes use of both host haloes and subhaloes. It specifies the relationship between the mean occupation number of galaxies and a certain halo/subhalo property, and the parameters are constrained through fitting both the 2PCFs and galaxy number density of a galaxy sample \citep{Guo16}. In our study, we use three halo/subhalo properties: \Macc, for a subhalo it is the mass at the redshift it ceased to be a distinct halo and for a distinct halo it is the present mass at $z=0$; \Vacc, similar to \Macc, but it is for the maximum circular velocity; \Vpeak, the peak of the maximum circular velocity over the history of a halo/subhalo. 

For each SCAM model, central and satellite galaxies occupy distinct haloes and subhaloes, respectively. The step-like functional form similar to equation~(\ref{eqn:hodcen}) (with halo mass replaced by the corresponding halo/subhalo property) is adopted to describe the central (satellite) mean occupation function in terms of the distinct halo (subhalo) population. There are two parameters for each mean occupation function, the characteristic quantity and transition width [the counterparts of $M_{\rm min}$ and $\sigma_{\log M}$ in eq.~(\ref{eqn:hodcen})]. That is, we have four parameters to describe the mean occupation functions. The traditional SHAM method can be regarded as a special case of the SCAM model. While SHAM assumes that satellites populate subhaloes in the same way as central galaxies populate distinct haloes, the SCAM model allows central and satellite galaxies to be different in occupying haloes/subhaloes, which is more flexible and likely more physical. For more discussions regarding the SHAM and SCAM  models and their relative performance in modelling the observational data, see \citet{Guo16}.

As with the HOD model, the two velocity bias parameters are also introduced in each SHAM model to model redshift-space clustering. We note that \Vpeak\ is the quantity used in HW13 to construct the mock galaxy catalogue through the SHAM method.
So we expect that the \Vpeak-based SCAM modelling result would closely reproduce the \hatfs\ curve from the HW13 mock. 
Among the four halo models, the HOD model does not account for any galaxy assembly bias effect, while any of the SCAM models has assembly bias built in through the halo assembly bias associated with the chosen halo property. Since halo assembly bias varies with halo properties \citep[e.g.][]{Gao07,Xu18}, it is interesting to see how well the \Macc- and \Vacc-based SCAM models work in reproducing the clustering measurements from the \Vpeak-based mock catalogue. The halo models we consider here therefore cover the cases ranging from zero assembly bias to various kinds of assembly bias. \par 
For the calculation of the galaxy 2PCFs, we adopt the accurate and efficient simulation-based method in \citet{Zheng16} using pre-compiled tables based on the Bolshoi haloes, binned in each halo/subhalo property. It is equivalent to populating haloes to create a mock catalogue for each set of model parameters and using the clustering measurements from the mock as the model prediction. 

For each model, we apply the MCMC method to explore the parameter space.
The likelihood is evaluated by the value of $\chi^2$ in the form of
\begin{equation}
\chi^2= \bmath{(\xi-\xi^*)^T C^{-1} (\xi-\xi^*)}
       +\frac{(\ngg-\ngg^*)^2}{\sigma_{\ngg}^2}, \label{eq:chi2}
\end{equation}
where $\ngg$ is the galaxy number density with an uncertainty $\sigma_{\ngg}$ (assumed to be $10\%$ of the measured value), the data vector $\bmath{\xi} = [\bmath{w_{\rm p}},\bmath{\xi_0},\bmath{\xi_2},\bmath{\xi_4}]$ includes the projected 2PCF $w_{\rm p}$, and the redshift-space multipoles $\xi_0$, $\xi_2$ and $\xi_4$, and  $\bmath{C}$ is the full error covariance matrix of the corresponding luminosity-threshold sample of the SDSS DR7 galaxies computed with the jackknife method. The quantity with (without) a superscript `$*$' represents the one from the measurement (model).

The best-fitting parameters for each model are then employed to create mock galaxy catalogues with the Bolshoi 
haloes/subhaloes. These mocks are then placed into redshift-space to compute the multipole moments and \hatfs, in
the same way as done with the original HW13 mock. 

For the HOD mocks, the number of central galaxy in a halo is either 0 or 1, drawn according to the probability in equation~(\ref{eqn:hodcen}), and the number of satellites is drawn from a Poisson distribution with the mean given by equation~(\ref{eqn:hodsat}). Each central galaxy is put at the centre of the corresponding halo (defined as the average position of the potential minimum; \citealt{Behroozi13a}). Its line-of-sight (l.o.s.) velocity $v_c$ with respect to that of the halo $v_h$, $v_c-v_h$, is drawn from a normal distribution with a standard deviation $\alpha_c\sigma_v$, where $\sigma_v$ is the one-dimensional (1D) velocity dispersion of the particles in the halo. Each satellite galaxy is assigned the position of a random particle in the halo,  with its l.o.s. velocity $v_s$ being that of the particle, $v_p$, modified by the satellite velocity bias factor, i.e. $v_s-v_h=\alpha_s (v_p-v_h)$. The SCAM mocks are similarly constructed but with the satellites being associated with random subhaloes. 

We note that the finite box size of the Bolshoi simulation would have little effect on our results. As we show later, the mocks to be compared are from the same simulation. Also our models themselves are built on the same simulation used to construct the mocks. That is, all our models adopt the method in \cite{Zheng16} and use the haloes/subhaloes in the Bolshoi simulation. The fluctuation powers on scales larger than the box size are missing in both the mocks and the models, or equivalently we study a universe that lacks those large-scale fluctuation powers. Therefore, it is legitimate to push the comparisons (e.g. on \hatfs) to substantially large scales (like tens of Mpc).

\begin{figure*}
\includegraphics[width=\textwidth]{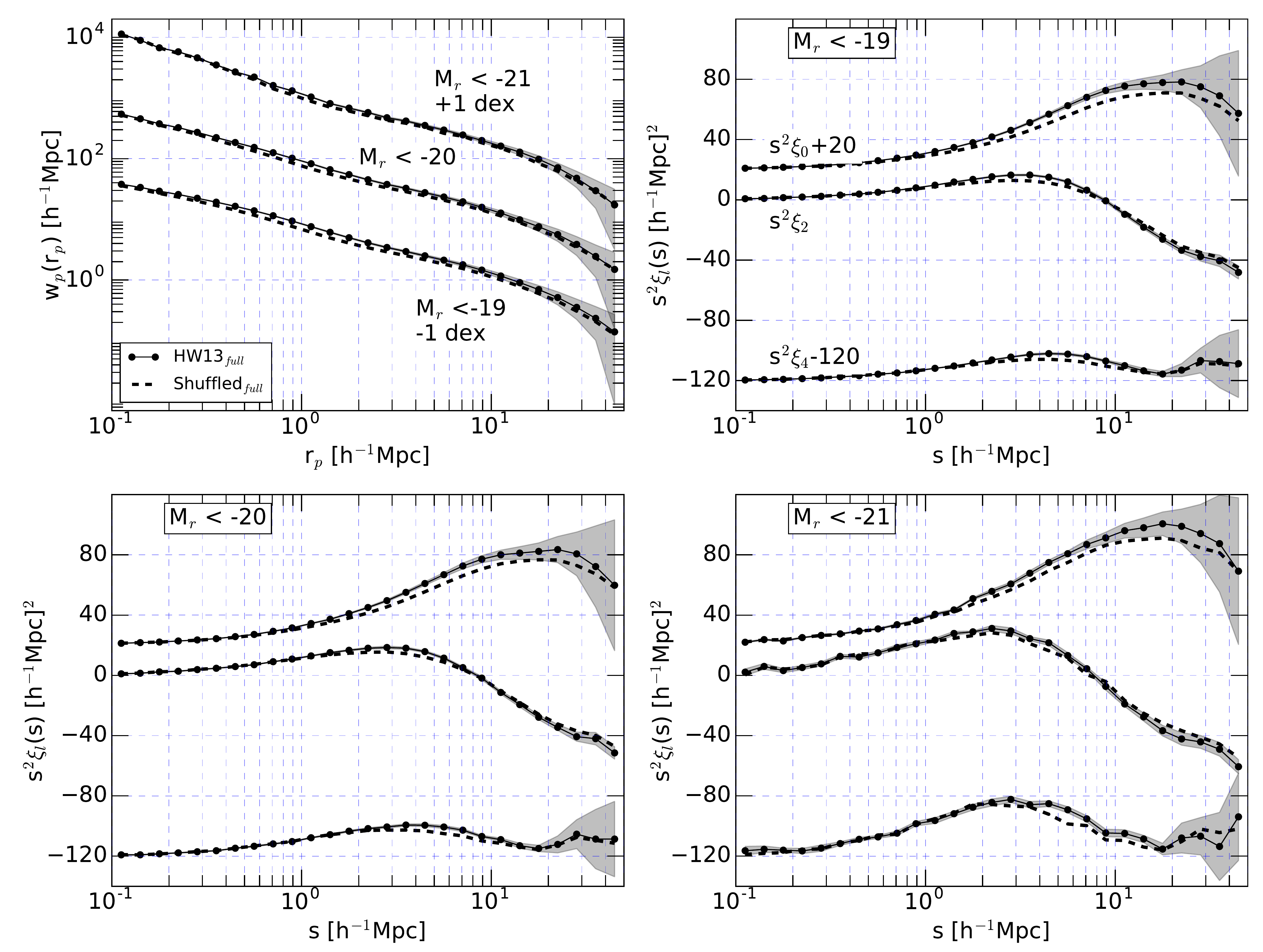}
\caption{
Comparison of projected 2PCFs and redshift-space multipoles from the HW13 mock and the shuffled mock.
Top-left: Comparison of the projected 2PCFs $w_{\rm p}(\rp)$ for three luminosity-threshold samples. 
For clarity, projected 2PCFs of the $M_r<-21$ and $M_r<-19$ sample have been shifted vertically up and down 
by 1 dex, respectively.
The other three panels show the multipoles for the three samples, and in each panel the monopole ($\xi_0$) and 
hexadecapole ($\xi_4$) curves have been shifted vertically for clarity. Jackknife errors are shown as shaded regions for the 
quantities associated with the HW13 mock.
}
\label{fig:mlt}
\end{figure*}
\begin{figure*}
\includegraphics[width=\textwidth]{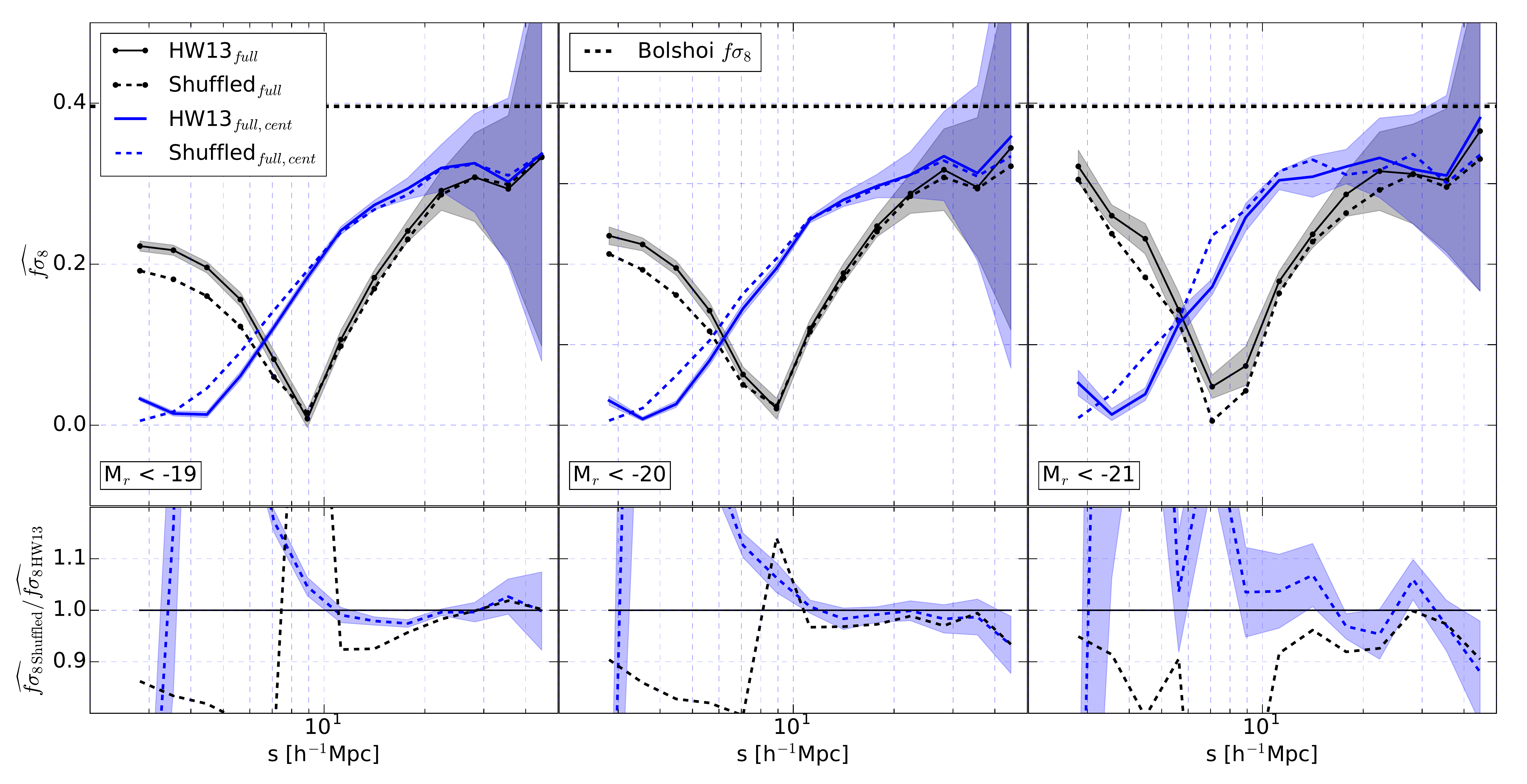}
\caption{
Comparison of the curves of the linear growth rate estimator \hatfs\ computed from the HW13 mock and the shuffled mock,
for three luminosity-threshold samples. In each top panel, the curves with (without) points are calculated based on all 
galaxies (central galaxies) in the sample, and solid (dashed) curves are for the HW13 (shuffled) mock. Jackknife
errors are shown for the curves associated with the HW13 mock. 
The horizontal dotted line marks the \fsig\ value
of the Bolshoi simulation that the mocks are based on.
In each bottom panel, the ratio of \hatfs\ computed from the shuffled mock and the HW13 mock from all galaxies and that from central galaxies are shown. For clarity, jackknife errors are only plotted for the latter.}
\label{fig:est_full}
\end{figure*}
\begin{figure*}
\includegraphics[width=\textwidth]{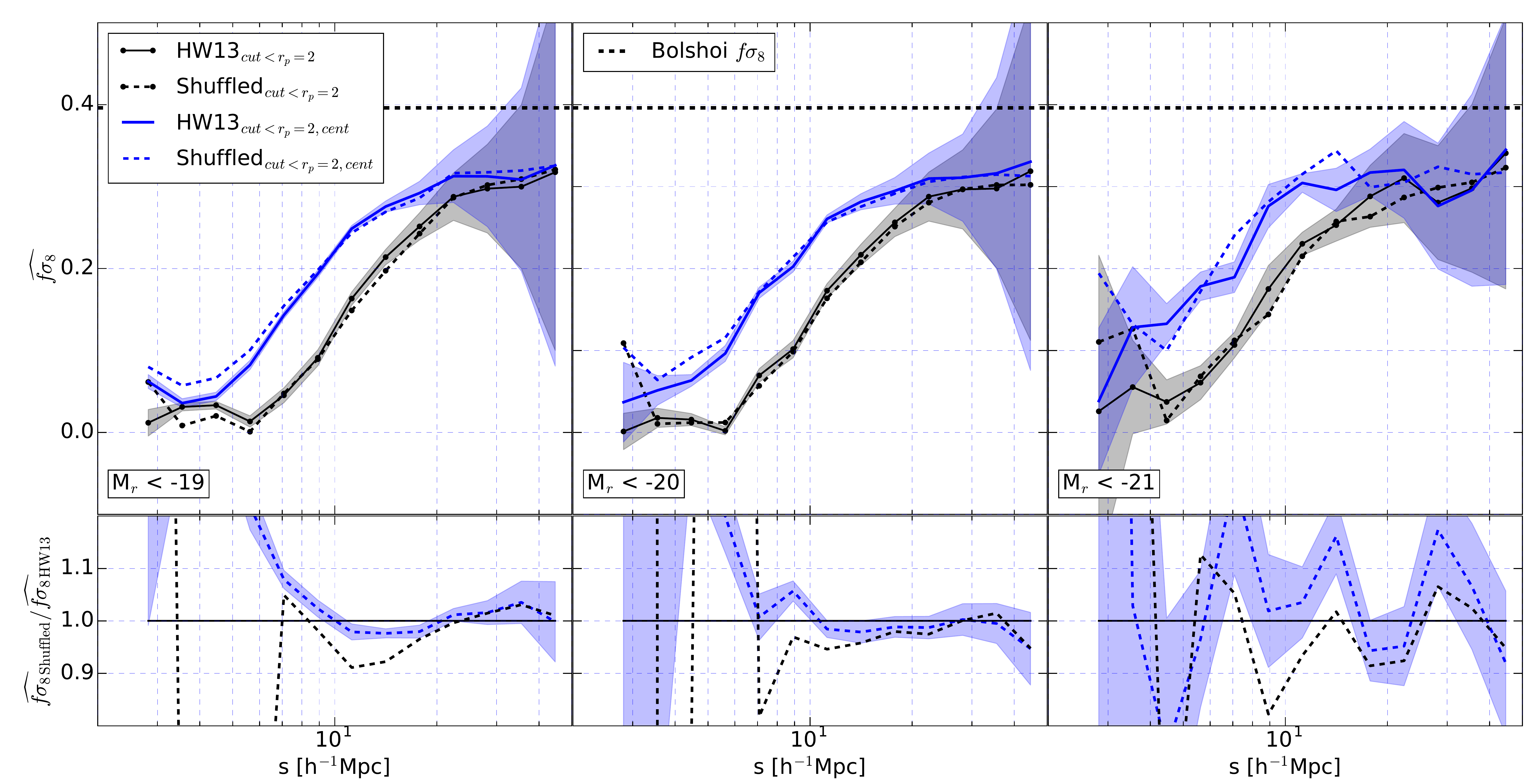}
\caption{
Similar to Fig.~\ref{fig:est_full}, but with multipoles calculated from the truncated redshift-space 2PCF.
The region with transverse pair separation $\rp<2\mpchi$ is excluded for calculating the multipoles, which 
mitigates the FOG effect. See text and Appendix~\ref{sec:B} for detail.}
\label{fig:est_cut}
\end{figure*}

\section{Results}\label{sec:Results}

\subsection{Galaxy clustering in the HW13 mock galaxy catalogue and the assembly bias effect}\label{sec:HW13result}

As a starting point of studying the assembly bias effect on the redshift-space clustering, we first compare the
clustering measurements with the HW13 mock catalogue (assembly bias included) and those with the shuffled catalogue
(large-scale assembly bias removed). 

Fig.~\ref{fig:rppi} shows the comparison in the redshift-space 2PCF $\xi(\rp,r_\pi)$ for the Mr19, Mr20, and Mr21 
samples. By design, the FOG parts of the original and shuffled samples are essentially the same, with the tiny difference caused by random motions of central galaxies close to each other \citep{Zheng16}. On large scales, the contours of the 2PCF of the shuffled sample appear to be more concentrated, being lower in amplitude at a fixed separation. That is, for the samples considered here, assembly bias leads to a more strongly clustered galaxy sample and a higher galaxy bias factor. 

The differences in the projected 2PCFs and redshift-space 2PCF multipoles are shown in Fig.~\ref{fig:mlt}. In general,
the (absolute) values of $w_{\rm p}$ and $\xi_{0,2,4}$ on most scales increase as the sample becomes more luminous. For 
each sample, on large scales, the projected 2PCFs from the mock with assembly bias show higher amplitudes, indicating
a higher galaxy bias factor. The higher bias factor also leads to higher (absolute) values of the multipoles. In the 
cases of monopoles $\xi_0$ and quadrupoles $\xi_2$, the difference can be understood with the Kaiser formula [e.g. 
equations~(\ref{eqn:modcorrarray1}) and (\ref{eqn:modcorrarray2})]. In the case of hexdecapoles $\xi_4$, the Kaiser formula
predicts a value independent of galaxy bias [e.g. equation~(\ref{eqn:modcorrarray3})]. In Fig.~\ref{fig:mlt}, we see that this is approximately true only on scales above $\sim20\mpchi$, indicating that the Kaiser formula becomes inaccurate 
below $\sim20\mpchi$.

In Fig.~\ref{fig:est_full}, we show the comparison between the \hatfs\ curves for the samples with assembly bias 
included and removed. The black curves are computed using the whole range of redshift-space 2PCFs. First, we notice
that even on the largest scales ($\sim 45\mpchi$) shown here, the value of the estimator \hatfs\ lies below the 
expected \fsig\ of the Bolshoi simulation (indicated by the thick dotted line). This is a manifestation of the 
inaccuracy of Kaiser formula on such scales, consistent with previous investigations \citep[e.g.][]{Reid11}. 
Given the construction of the estimator and the application scales of the Kaiser formula, we do not expect the 
\hatfs\ curve on intermediate and small scales to have the exact value of \fsig, either. For a given galaxy sample, it is the 
whole \hatfs\ curve that encodes the \fsig\ information, and an RSD model that fits the curve would lead to 
constraints of \fsig. Here we use the \hatfs\ curves to serve our purpose of comparison to study the assembly bias effect. We find that the solid and dashed black curves track each other well on scales above 9$\mpchi$.\footnote{We choose a difference of 5\% as an approximate tolerance threshold to define the scales. Note that the Mr21 sample is noisier, as it has the lowest number of galaxies in the volume. The scales we quote are mainly from the Mr19 and the M20 samples, which also roughly applies to the Mr21 sample given the uncertainty.} 

It may suggest that the assembly bias effect would not influence \hatfs\ down to 9$\mpchi$. However, we note that the \hatfs\ curve has a substantial contribution from the FOG effect, which causes \hatfs\ to cross zero around 9$\mpchi$. Since the FOG comes from random motions of galaxies inside haloes, there is not much cosmological information to extract. In fact, most of the cosmological information (like \fsig) is encoded in the halo velocity field, which tracks the matter velocity field and is determined by the structure growth rate. It then makes sense to compare the \hatfs\ curves determined from central galaxies (blue solid and dashed curves) whose motion is largely related to the halo velocity field. It is encouraging that the two curves closely track each other down to $\sim 9\mpchi$, which suggests that the RSD on such scales can still be used to contribute to the \fsig\ constraints even with the existence of assembly bias effect as strong as in the HW13 mock.

\begin{figure*}
\includegraphics[width=\textwidth]{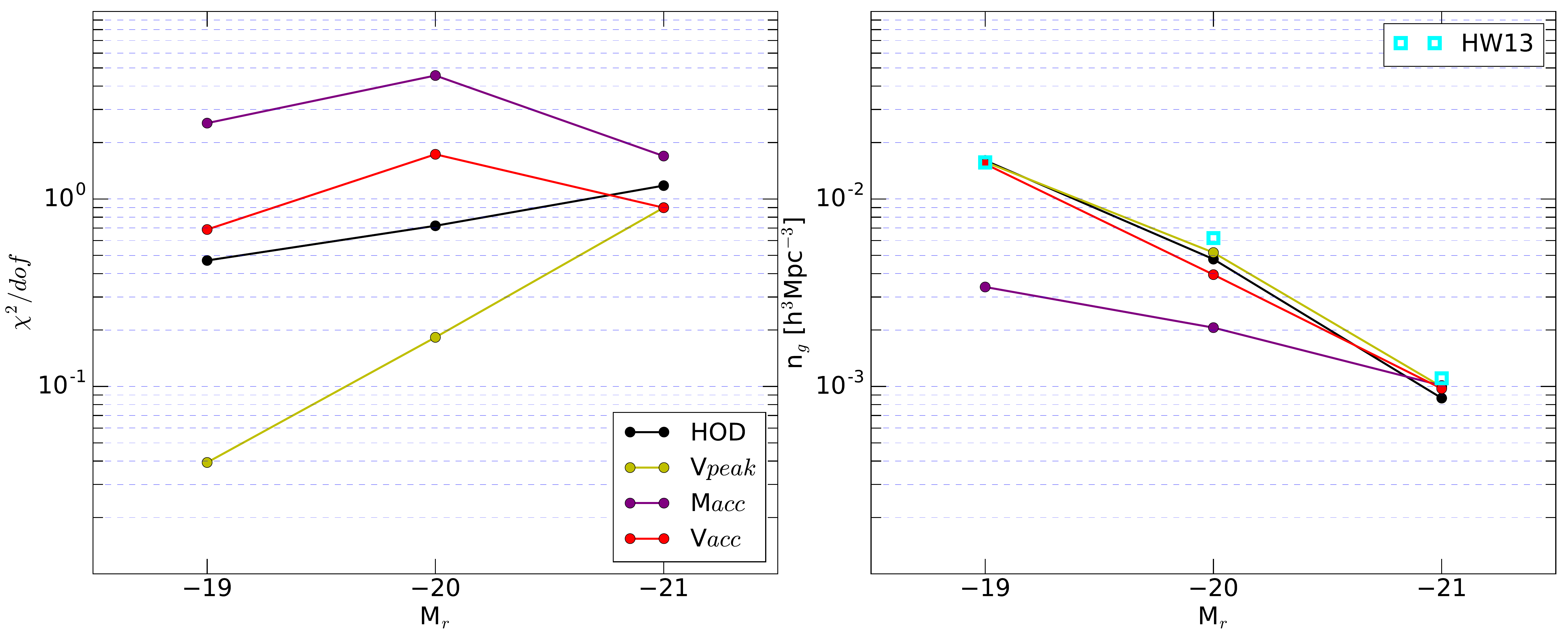}
\caption{Left: Reduced $\chi^2$ values for the best-fitting halo models of the three luminosity-threshold samples from the HW13 mock. The values of $\chi^2$ are calculated according to Eq.~\ref{eq:chi2}, with the degrees of freedom being $42$ for the HOD model and $43$ for the SCAM model. Right: Best-fitting number densities in comparison with those from the HW13 mock, the differences of which contribute to the $\chi^2$ values.}
\label{fig:chi2}
\end{figure*}

In reality, it is not easy to construct a central-only galaxy sample, and it is preferable to have the model separate the
contributions of central and satellite galaxies. In our exercise here, we can devise a way to mitigate the small-scale
FOG effect by excluding the 2PCF data measured at small transverse separation $\rp$ and obtaining the multipoles through
combinations of the modified multipoles (see Appendix~\ref{sec:B} for detail). In Fig.~\ref{fig:est_cut}, we show the \hatfs\ curves from the redshift-space 2PCF that excludes the part with $\rp<2\mpchi$. For the case of central galaxies only (blue curves), the curves with assembly bias included/removed track each other down to $\sim 9\mpchi$, similar to the case without the truncation in the data (blue curves in Fig.~\ref{fig:est_full}). For the cases with all the galaxies (non-blue curves), the smoothing caused by two-halo central-satellite galaxy pairs helps to drive the matching scale down to $\sim 6\mpchi$. 

Overall the comparisons between the results with the assembly bias included/removed suggest that the 
intermediate-scale RSD data can contribute to tighten the \fsig\ constraints even with the existence of
assembly bias. With assembly bias as strong as in the HW13 catalogue, the RSD measurements on scales as
small as $\sim 9\mpchi$ can be used for the above purpose, based on the comparison with central galaxies. While the case 
with all galaxies indicates even smaller scales, the information is likely entangled by galaxy kinematics inside haloes. 
Since the \hatfs\ curve with the truncated 2PCF is what we derive with the FOG effect mitigated, in what follows, we 
will present the results based on the truncated data and then use the central-only case to guide the interpretation.

\begin{figure*}
\includegraphics[width=\textwidth]{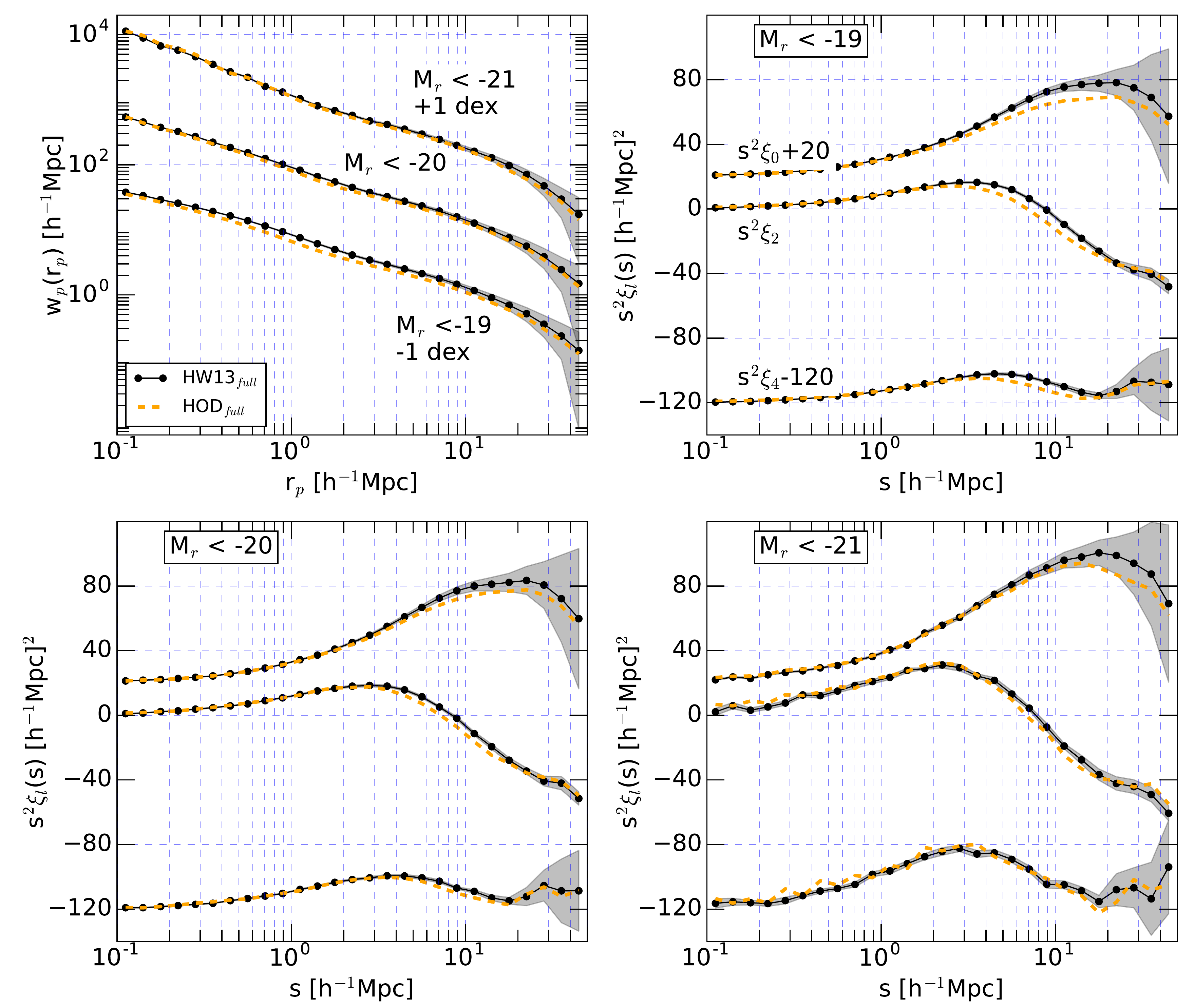}
\caption{
Same as Fig.~\ref{fig:mlt}, but comparing the 2PCFs and multipoles from the HW13 mock and those from the best-fitting
HOD model.}
\label{fig:mlthod}
\end{figure*}
\begin{figure*}
\includegraphics[width=\textwidth]{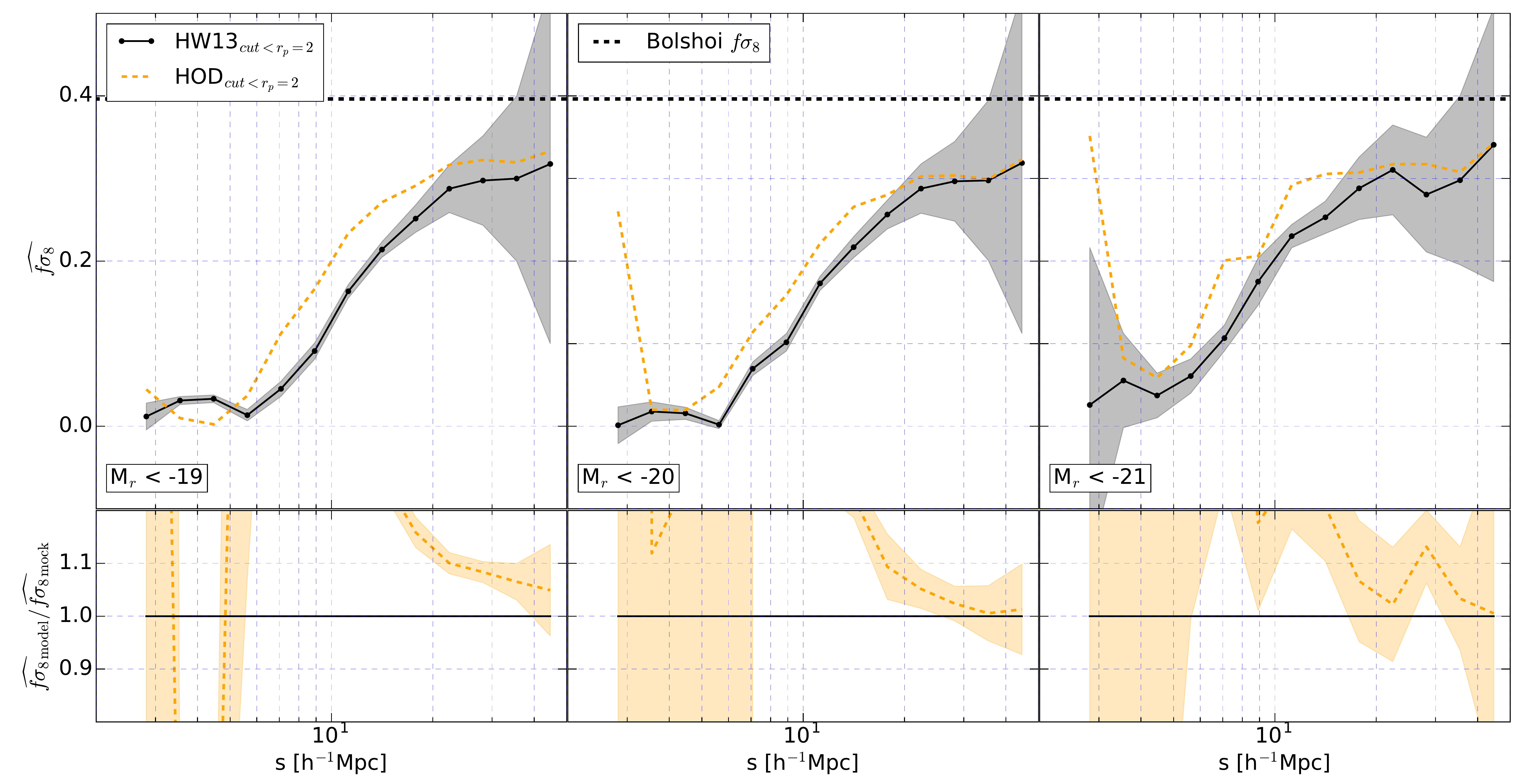}
\caption{
Similar to Fig.~\ref{fig:est_cut}, but comparing the \hatfs\ curves from the HW13 mock and those from the best-fitting
HOD model. In each bottom panel, the ratio of the two curves is shown, where the uncertainty in the ratio is calculated with the jackknife method.}
\label{fig:esthodcut}
\end{figure*}

\begin{figure*}
\includegraphics[width=\textwidth]{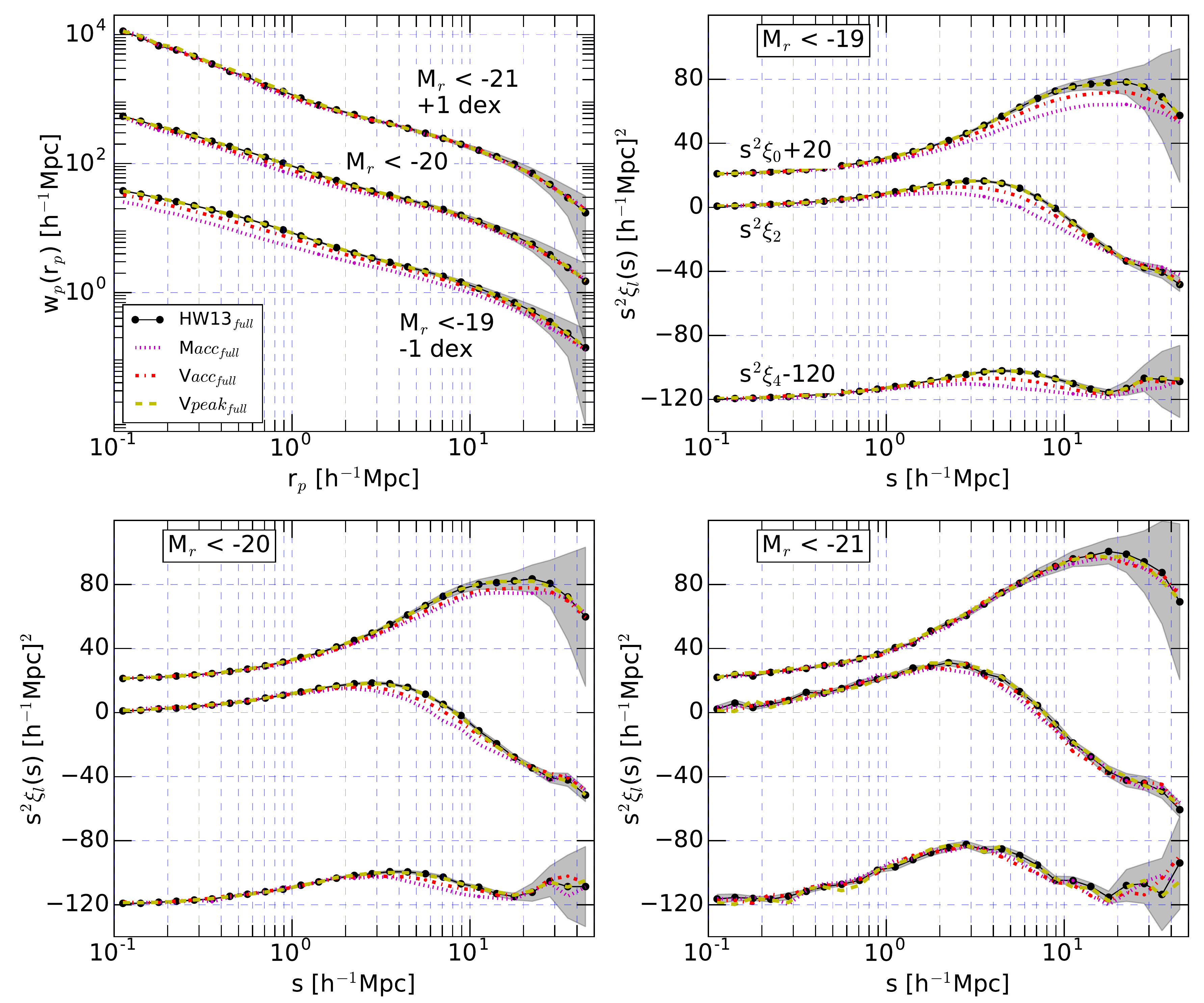}
\caption{
Same as Fig.~\ref{fig:mlthod}, but comparing the 2PCFs and multipoles from the HW13 mock and those from the best-fitting
SCAM model.}
\label{fig:mltscam}
\end{figure*}

\begin{figure*}
\includegraphics[width=\textwidth]{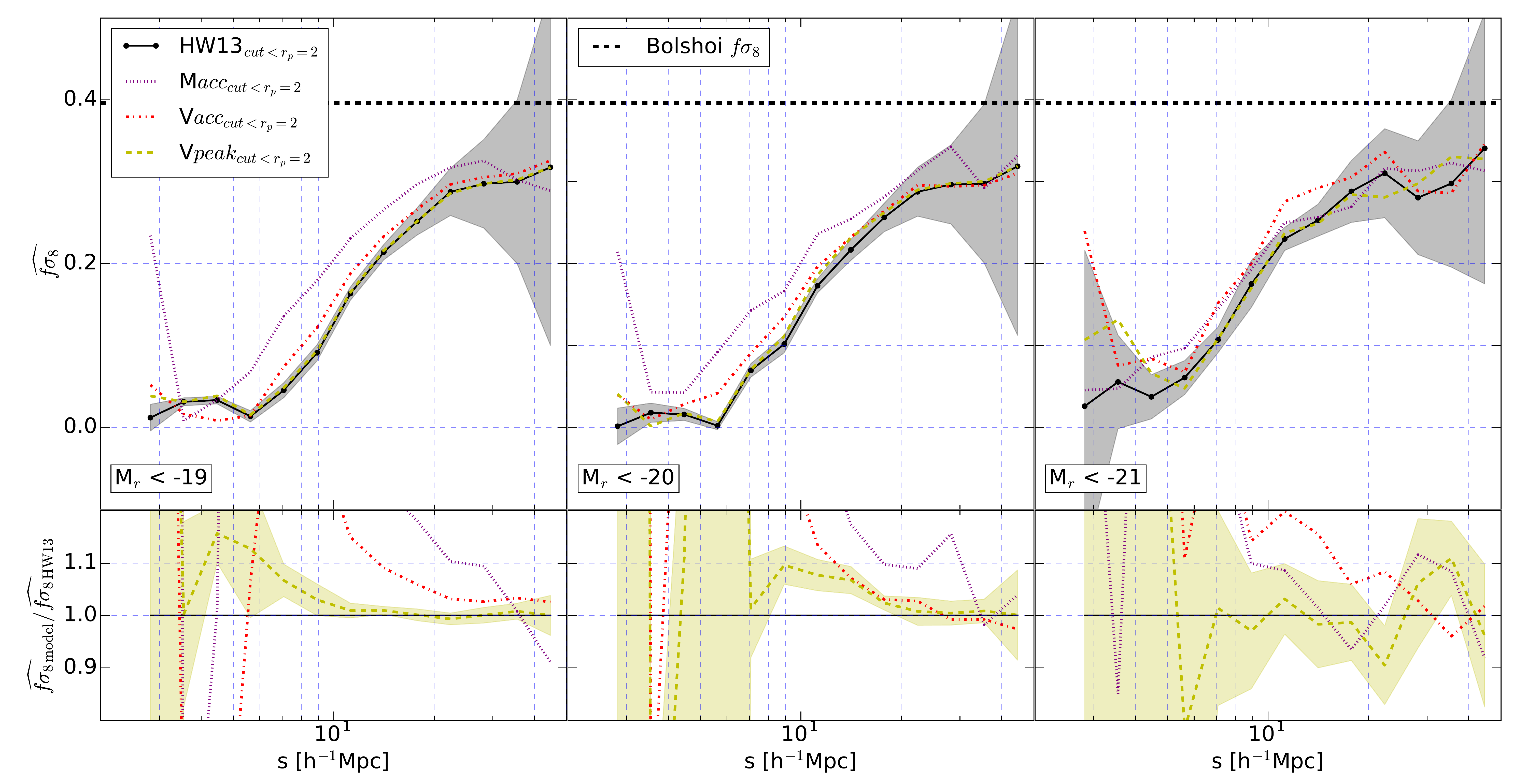}
\caption{
Similar to Fig.~\ref{fig:esthodcut}, but comparing the \hatfs\ curves from the HW13 mock and those from the best-fitting
SCAM models. 
In each of the bottom panel, for clarity, only the uncertainty from the jackknife method for the \Vpeak\ model is shown, and those for the other models are similar in magnitude.}
\label{fig:estscamcut}
\end{figure*} 

\subsection{Halo modelling results}

In this subsection, we present the halo modelling results, including those from the HOD and the three SCAM models. 
The values of $\chi^2$ [eq.~(\ref{eq:chi2})] per degrees of freedom ({\it dof}; 
42 for the HOD model and 43 for each SCAM model) from the best-fitting models and the best-fitting number densities from the models are shown in Fig.~\ref{fig:chi2}. In general, the \Vpeak\ model achieves the lowest $\chi^2$, which is not surprising as the HW13 mock is \Vpeak-based, and the HOD and \Vacc\ models have reasonable values of $\chi^2$ (left panel). These three models are able to reproduce the number densities for the Mr19 and Mr21 samples but predict  slightly lower values for the Mr20 sample (right panel). The \Macc\ model does not provide good fits to the data ($\chi^2/dof\ga 2$), and the predicted number densities for the Mr19 and Mr20 samples are significantly low in compensation for fitting the 2PCFs (see \citealt{Guo16} for a similar effect).  We will continue discussing Fig.~\ref{fig:chi2} in the subsections below.

\subsubsection{HOD modelling results}\label{sec:HODresults}

The standard HOD framework describes the occupation statistics of galaxies as a function of halo mass, not any 
assembly properties. If the assembly bias of haloes is inherited in any significant way by the galaxies, the HOD
modelling may give incorrect or biased inference on the galaxy-halo connection and cosmology 
\citep[e.g.][]{Zentner14}. \citet{McEwen16} find that the standard HOD and an extended version that accounts for 
halo environment can both describe the galaxy-matter correlation well enough to reproduce the matter correlation
function beyond $\sim 2\mpchi$, based on an estimator independent of galaxy bias. It indicates that an extension 
to the HOD model to include the assembly bias effect is not necessary for inferring the matter correlation function
from galaxy clustering and galaxy lensing.  Motivated by the results in \S~\ref{sec:HW13result}, here we investigate 
whether the HOD model with no assembly bias is sufficient to be used to constrain \fsig\ with small- and 
intermediate-scale RSD data. In the next subsection we will discuss the results with the SCAM models, which are our 
models with `environmental' dependence and have certain assembly bias effect built in.

With the HOD model, the best-fitting results to the projected 2PCFs and redshift-space 2PCF multipoles are shown in 
Fig.~\ref{fig:mlthod}. For the projected 2PCF, the HOD best-fitting results match those from the HW13 mock, especially 
for the two most luminous samples. This is consistent with previous results that the HW13 galaxy mock reasonably reproduces the projected 2PCFs of the SDSS DR7 galaxies \citep{Hearin13} and that the HOD framework is also able to model those well \citep[e.g.][]{Guo15b,Guo16}. For each of the three luminosity-threshold samples, the 
HOD model fits the redshift-space multipoles on small scales, while it under predicts those on large scales. 
Interestingly the differences between the HW13 2PCFs and the HOD fits are similar in trend to those seen in 
the comparison between mocks with assembly bias included and removed (see Fig.~\ref{fig:mlt}), a result not
unexpected. It seems that the HOD model is unable to successfully interpret the redshift-space galaxy clustering amplified by the assembly bias effect in the HW13 construction. However, we note that the values of $\chi^2$ from the
best-fitting HOD models are reasonable (see Fig.~\ref{fig:chi2}), $19.73$, $30.21$, and $49.48$, for the three samples 
(with 42 degrees of freedom for each sample), respectively, owing to the covariances of data points on large scales 
(e.g. fig.~3 of \citealt{Guo16}).  
The number densities of the three samples are also reasonably reproduced (right panel of Fig.~\ref{fig:mlt}).

In Fig.~\ref{fig:esthodcut}, we compare the \hatfs\ curve from the best-fitting HOD model of each sample and 
that from the HW13 mock,
with the multipoles calculated from truncated 2PCFs. The HOD \hatfs\ curve only approaches the HW13 on scales above
20$\mpchi$ and deviates from it towards small scales, with the difference reaching tens of per cent around 10$\mpchi$. 
That is, HOD modelling without accounting for assembly bias fails to recover the expected \hatfs\ on scales of
$\sim 10 \mpchi$. The result seems to imply that the HOD model could not enable us to use the RSD on such scales to 
tighten \fsig\ constraints. We will return to this discussion after presenting the results with SCAM modelling.

\subsubsection{SCAM modelling results}\label{sec:SCAMresults}

The SCAM models are based on halo/subhalo properties with assembly effect encoded, and those models serve as our halo 
models with assembly bias effect included to some extent. Of the three SCAM models we consider, the one based on
the peak maximum circular velocity \Vpeak\ should fully capture the assembly bias effect in the HW13 mock, given that 
the construction of the mock is based on \Vpeak. For the model with maximum circular velocity \Vacc\ at the time of
accretion, the assembly bias effect differs from the \Vpeak\ model. For the model with mass \Macc\ at the time of 
accretion, as it uses halo masses for distinct haloes, there would be no two-halo assembly effect and its main 
difference from the HOD model is the distribution of satellites inside host haloes. The SCAM models considered here 
therefore can cover a range of assembly bias effect.

In Fig.~\ref{fig:mltscam}, we see that the \Vpeak\ model nearly perfectly reproduces the HW13 clustering measurements (see the low $\chi^2/dof$ in Fig.~\ref{fig:chi2}), as expected. The \Vacc\ model fits the data reasonably well, with $\chi^2/dof$ near unity for the Mr19 and Mr21 sample and about two for the Mr20 sample. The \Macc\ model is not able to provide a good match to the data, with $\chi^2/dof \sim$ 2.5, 5, and 2, for the Mr19, Mr20, and Mr21 sample, respectively. It predicts much lower galaxy number densities for the Mr19 and Mr20 samples (right panel of Fig.~\ref{fig:chi2}), 
with a trend of progressively lower for samples of higher satellite fractions.
As mentioned before, compared to the HOD model, the \Macc\ model is different in the distribution of satellites, and thus it is the satellite occupation that drives the \Macc\ model to behave much worse than the HOD model. 
The results with the \Macc\ model is similar to those found in fitting redshift-space clustering of SDSS DR7 galaxies \citep{Guo16}. As in \citet{Guo16}, it is mainly the quadrupole that is unable to be reproduced by the \Macc\ model. It indicates a difference in the velocity distributions of the \Macc\ subhaloes and DM particles (used in the HOD model), and even the velocity bias is not able to correct the difference and bring \Macc\ and HOD models into agreement. The \Macc\ model twiddles between the fits to the 2PCFs and number density to achieve the overall best fit, and the lower best-fitting number density is a result of the compromise.
As a whole, for the SCAM models, the assembly bias in quantities other than \Vpeak\ could not fully capture the effect encoded in \Vpeak\ to reproduce the redshift-space clustering measurements from the \Vpeak-based mock. \par 
The \hatfs\ curves from the best-fitting SCAM models are compared in Fig.~\ref{fig:estscamcut}. Besides the perfect match with the \Vpeak\ model, the predictions from the \Vacc\ and \Macc\ model deviate from the expected curve, and the deviation increases towards small scales. The deviations depend on the sample, decreasing for more luminous samples. For example, at 8$\mpchi$, the fractional difference goes from $\sim$50, $\sim$30, to $\sim$25 per cent for the Mr19, Mr20, and Mr21 samples with the \Vacc\ model. We seem to reach a conclusion similar to the HOD modelling case -- if we do not know the halo property that the galaxy assembly bias most closely ties to, we would not be able to use RSD on scales below $\sim 20\mpchi$ to help constrain \fsig. However, this may not be true, as we discuss in the next subsection. \par 

\subsection{Further clues from centrals-only results and SDSS DR7 measurements}
 
The halo modelling results shown in the above subsections seem to demonstrate that the assembly bias effect,
if not correctly accounted for, would only allow us to use RSD on scales above 20$\mpchi$ to constrain \fsig.
However, there are two considerations that may lead us to circumvent the apparent results. First, although we
avoid the FOG effect in defining the \hatfs\ curve with modified multipoles from truncated data, we still have 
contributions from satellites (in the two-halo regime). Since the cosmological information is in the halo
velocity field, which is more directly probed by central galaxies, it would be necessary to check the \hatfs\ curve
from central galaxies to see whether the above apparent conclusion still holds. Second, our analyses assume that
the galaxy assembly bias is as strong as that in the HW13 catalogue. We would like to see whether the 
assembly bias in reality can be weaker, and further comparisons between HW13 mock and SDSS data would be useful.

\begin{figure*}
\includegraphics[width=\textwidth]{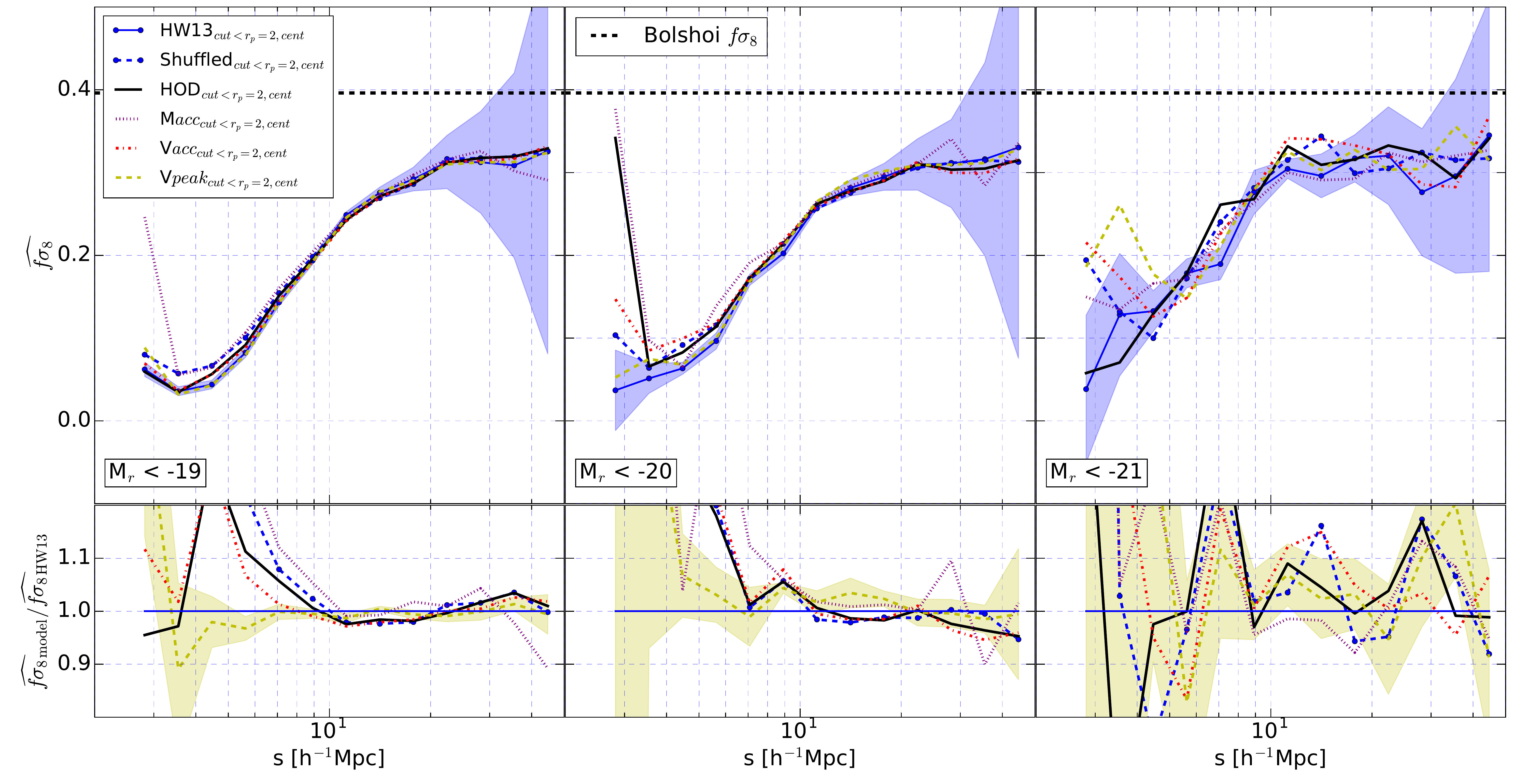}
\caption{
Similar to Fig.~\ref{fig:estscamcut}, comparing the \hatfs\ curves using only central galaxies from the HW13 mock and 
those from the best-fitting HOD and SCAM models.}
\label{fig:estcentcut}
\end{figure*}

Fig.~\ref{fig:estcentcut} compares the \hatfs\ curves from the best-fitting HOD and SCAM models with that measured
from the HW13 mock, all computed using only central galaxies and the truncated data. Interestingly the curves from
best-fitting halo models show agreement with the HW13 measurements down to scales around 8$\mpchi$. Therefore, the halo 
models considered here are able to describe the \hatfs\ curve probed by central galaxies down to scales of 8$\mpchi$,
and there is hope to extract the \fsig\ information with data on such scales as it is encoded in the kinematics of
distinct haloes with central galaxies. As the key difference here from the \hatfs\ results in 
sections~\ref{sec:HODresults} and \ref{sec:SCAMresults} is that the satellites are removed, the disagreements seen in 
the Fig.s~\ref{fig:esthodcut} and \ref{fig:estscamcut} are caused by the distribution of satellites. If in each halo 
model we change the prescription for satellite occupation distribution and add more flexibility to it, it is possible 
to reach better fits to the measurements. As the growth rate information is in the central galaxies, the satellite
occupation distributions in the model serves as nuisance parameters to be marginalised. Since in practice it is 
impossible to have a pure and complete sample of central galaxies, a halo model with flexible satellite prescriptions
would probably be the way to model the RSD on small scales and to single out the central galaxy contribution to 
constrain \fsig. While the constraining power is better investigated with such a model, our finding here based on
the behaviour of \hatfs\ of central galaxies is encouraging, and a model with no assembly bias or with assembly 
bias different from reality could still be used to extract cosmological information from the RSD data on small and intermediate scales.

\begin{figure*}
\includegraphics[width=\textwidth]{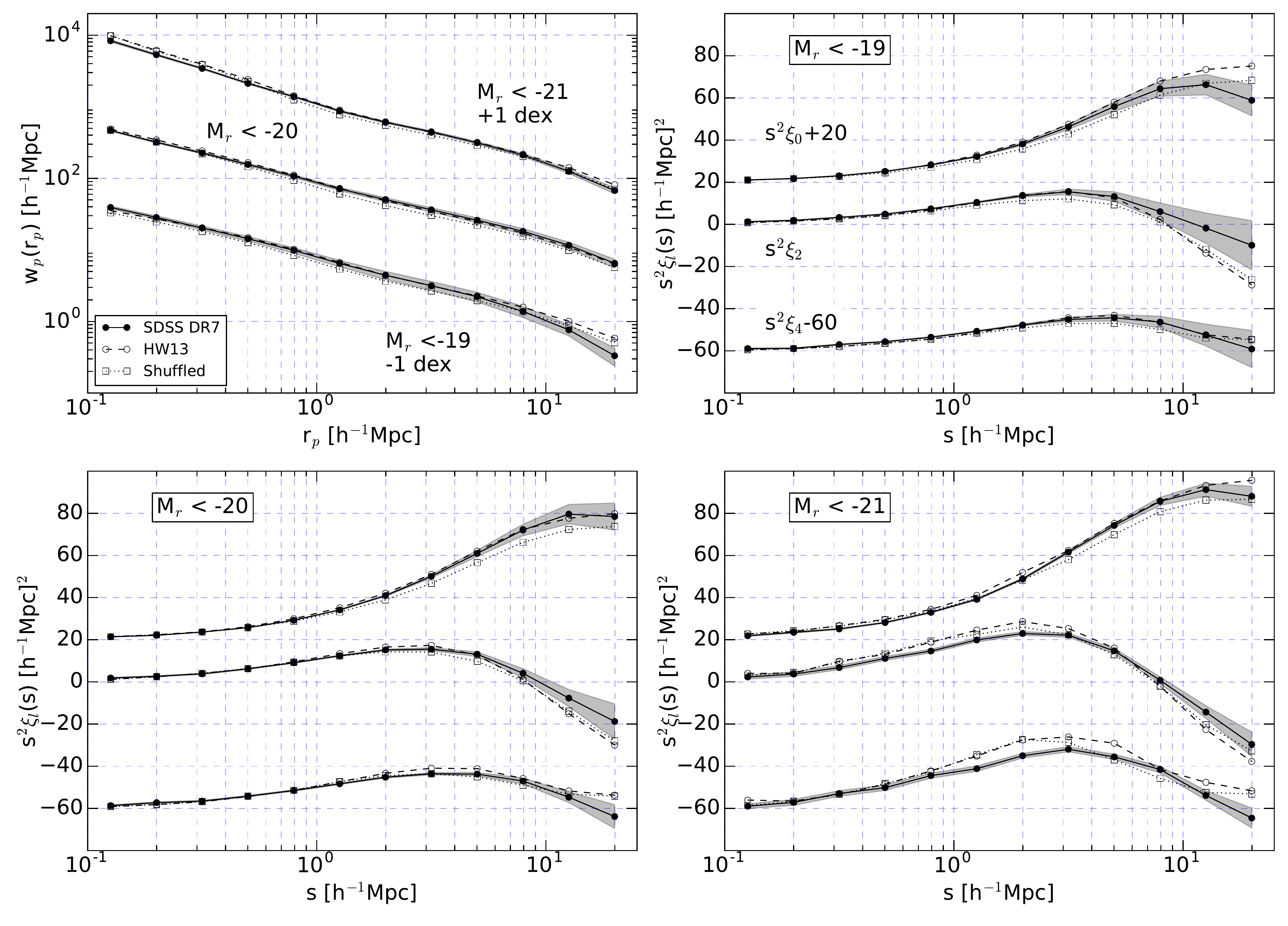}
\caption{
Similar to Fig.~\ref{fig:mlt}, but comparing the 2PCFs and multipoles from the HW13and the shuffled mocks and those from 
the SDSS DR7 data.}
\label{fig:mltdr7}
\end{figure*}

The assembly bias in the HW13 mock comes from relating galaxy luminosity to \Vpeak, which has halo assembly bias. 
Although the mock can reasonably reproduce the projected 2PCFs from the SDSS DR7 data, it does not necessarily mean that the assembly bias in the real universe is similar. 
We can test this by extending the comparison between HW13 mock predictions and SDSS DR7 measurements to other clustering statistics. 
In Fig.~\ref{fig:mltdr7}, in addition to the projected 2PCFs, we compare the redshift-space 2PCF multipoles measured from the HW13 and the shuffled mocks and those from the SDSS DR7 data for the three luminosity-threshold samples. For each sample, there are clear and significant deviations of the HW13 and the shuffled measurements from the SDSS DR7 results. 
For a quantitative assessment, we compute the values of $\chi^2$ with the mock predicted and SDSS measured 2PCFs and obtain $163$, $184$, and $316$ with the three HW13 mock samples and $158$, $463$, and $368$ with the three shuffled samples, each with 36 degrees of freedom. 
Although the HW13 mock can match the projected 2PCFs reasonably well, it does not lead to good fits to the redshift-space clustering, implying that the assembly bias encoded in the HW13 mock may not be realistic 
(see Appendix~\ref{sec:C} for the case with galaxy samples defined by colours).
The shuffled mock is no better at matching the SDSS measurements, even showing a worse prediction for the Mr20 and Mr21 samples. This seems to imply that the redshift-space clustering is of no help to constrain assembly bias. However, we note that the shuffled mock is not completely free of assembly bias signal (as we keep the 1-halo configuration of the HW13 mock). \citet{Guo16} demonstrates that the HOD model (free of assembly bias) can well fit the SDSS redshift-space measurements, with a better performance than the \Vpeak\ SCAM model for the Mr19 and Mr20 samples (see their fig.13). The success of the HOD model in interpreting the redshift-space clustering by no means rules out the presence of galaxy assembly bias in the real data, but whatever form the assembly bias may be it could be different from that in the \Vpeak\ SCAM model or the HW13 mock. For a potentially general form of galaxy assembly bias, the constraining power of the redshift-space clustering on it remains as a topic for further investigations.

For the luminosity-threshold samples, the galaxy assembly bias in the HW13 mock comes from linking galaxy luminosity with halo/subhalo \Vpeak. The correlation between the two quantities may not be as strong as assumed in HW13. It would be useful to  examine such a correlation in galaxy formation models, which is expected to depend on the implementation of baryon processes.

Observational study of galaxy assembly bias would also be useful and complementary. With observations, it is still under investigations whether galaxy assembly bias exists or how strong it is. For example,
\citet{Lin16} find no evidence of galaxy assembly bias by studying the clustering of early and late central galaxies
with the host halo mass controlled by galaxy lensing measurement. If galaxy assembly bias is weak as indicated, it 
would make the halo models (e.g. HOD) in an even better position to use the small-scale RSD data to tighten constraints
on \fsig.


\section{Conclusion and Discussion}\label{sec:discussion}

The RSD effect has been used to learn about the cosmic linear growth rate \fsig, which can help probe the nature of
the accelerated expansion of the universe by constraining parameters of dark energy or testing theories of gravity.
Usually the \fsig\ constraints come from RSD measurements on large scales, where the model is relatively simple
and we do no need to worry too much about galaxy formation physics. Because of the high statistical power of small- 
and intermediate-scale RSD, extending the RSD analysis towards small scales in principle can help tighten the
constraints on \fsig. However, galaxy assembly bias, if not correctly modelled, may prevent us from using the
small-scale RSD data to extract cosmological information. In this paper, using a mock galaxy catalogue with 
built-in assembly bias, we perform a preliminary study on how assembly bias may affect the inference of \fsig\ and
whether commonly adopted halo models are able to bypass the assembly bias effect for \fsig\ constraints.

Our study is motivated by and to some extent in parallel to the investigation by \citet{McEwen16}. They find that 
commonly adopted halo models can sufficiently describe the small-scale galaxy-matter correlation coefficient and that
one can infer the correct matter correlation function down to scales of a few $\mpchi$ with galaxy clustering and
weak lensing data even if assembly bias is not correctly modelled. By extending the \fsig\ estimator in 
\citet{Percival09} to configuration space, we devise an estimator \hatfs\ based 
on redshift-space multipoles to represent the \fsig\ information on both small and large scales in the RSD data. 
A comparison between 
\hatfs\ curves from the catalogues with and without assembly bias shows that the two curves overlap on scales 
down to 8--9$\mpchi$, implying that even with assembly bias as strong as in the HW13 catalogue we can still 
expect to extract \fsig\ information from such scales. 

We then apply the HOD model (with no assembly bias incorporated) and three SCAM model (based on \Vpeak, \Vacc, and
\Macc, with various forms of assembly bias) to fit the projected 2PCF and redshift-space multipoles from the HW13
mock catalogue with assembly bias. By design, the \Vpeak\ SCAM model reproduces the measurements, as the mock is 
constructed through the SHAM method using \Vpeak. The other three models, however, could not produce good fits to 
the mock measurements, and the \hatfs\ curves deviate substantially from the expected one on scales below 
$\sim 20\mpchi$. This seems to indicate that without knowing the origin of assembly bias in the galaxy sample, the halo
model would not help constrain \fsig\ with the RSD data on small scales. However, when turning to \hatfs\ curves
determined from central galaxies, we find that all models are able to match the curve from the HW13 mock down to 8$\mpchi$. As the 
\fsig\ information is encoded in the motion of haloes, probed by that of the central galaxies, the result shows that
the intermediate-scale RSD data could still be used for constraining \fsig. The failure of the HOD and the \Vacc/\Macc\
SCAM models in reproducing the mock measurements with all galaxies lies in the insufficient description of 
the occupation of satellite galaxies.

Our results suggests that if we could properly identify central galaxies in a galaxy sample, we could then utilise the HOD or SCAM models to model the RSD down to scales of $\sim 8\mpchi$ without concerning that systematic errors are introduced by not properly accounting for the galaxy assembly bias. However, identifying central galaxies in a galaxy sample (e.g. through identifying galaxy groups; \citealt{Yang05}) is not straightforward and can hardly reach the level of high purity and completeness desired for the application. As central galaxies and satellite galaxies are separated in the HOD or SCAM halo model, it would be desirable to separate out central galaxies through the model instead of constructing a sample of central galaxies. We see that even though the satellite phase-space and occupation distribution may not be accurate, the model can still capture the correct RSD signal from central galaxies, which contains the cosmological information. To improve the model, it is necessary to make the prescription for satellites more flexible so that it can describe the small-scale FOG effect well enough to remove the negative impact on accurately extracting the clustering of central galaxies. The components that could be modified in the HOD model include the scatter in the satellite occupation number \citep[e.g.][]{Dvornik18}, the spatial distribution of satellites inside haloes, and the velocity bias of satellites \citep[e.g.][]{Guo15a}.

The effect of assembly bias on redshift-space clustering was investigated before based on mock galaxy catalogues from semi-analytic galaxy formation models and the shuffled control samples \citep[][]{Zu08,Padilla18}. For the effect on cosmological constraints, both investigations use the large-scale RSD, informed by $\beta=f/b$ from the ratio of monopole to real-space 2PCF and/or the quadrupole. As point out in \citet{Padilla18} and \citet{Xu18}, assembly bias affects the large-scale spatial clustering and velocity field of galaxies/haloes consistently, leading to little effect on cosmological constraints in \fsig. Different from those studies, we focus our investigation on the effect of assembly bias on the intermediate- and small-scale RSD in the hope that the \fsig\ constraints could benefit from the high statistical power of clustering measurements on such scales. 

Our results in this paper are based on comparisons with the HW13 mock catalogue, which introduces galaxy assembly bias by associating galaxy properties with \Vpeak. Galaxy formation models may predict different forms and degrees of assembly bias. It would be useful to carry out similar analyses with mock catalogues from semi-analytic galaxy formation models and hydrodynamic simulations to see how sensitive our results are to different assembly bias signals. Given that \Vpeak\ and related quantities are found to capture a large extent of galaxy assembly bias in hydrodynamic simulations \citep[e.g.][]{Chaves-Montero16,Matthee17,Xu19}, we speculate that our results would not have a substantial change. In addition, the assembly bias effect in real data may be weaker than that in the HW13 catalogue \citep[e.g.][]{Lin16,Guo16}, implying that our results are likely conservative. Nevertheless, without a full investigation using different galaxy formation models, our results, especially the numbers (such as the minimum scales), should be taken as a broad estimate.

\citet{McEwen16} conclude that the matter correlation function can be accurately recovered down to scales of $\sim 4 \mpchi$ even without an accurate assembly bias model. Here our investigation indicates that it is possible to use the RSD down to scales of $\sim 8 \mpchi$ to constrain \fsig\ even if we lack the knowledge of galaxy assembly bias. Although it is an encouraging message, our study is still preliminary. First, we focus on a scale-dependent estimator \hatfs\ as a convenient metric that encodes the \fsig\ information, not the \fsig\ constraint itself from a full model. We could not tell how the deviations from the expected \fsig\ curve translate to the systematic errors in the \fsig\ constraints. Second, our results are more qualitative than quantitative. For example, the information contents in the RSD at various scales for \fsig\ constraints cannot be addressed with our preliminary investigation, and we do not know the relative contributions from different scales. Furthermore,  it is not clear how the cosmological information is degenerate with the (central) galaxy velocity bias. 

A further and thorough investigation as our future work is to apply a full model of the RSD to the measurements from mocks built with various forms and degrees of assembly bias (besides that ingrained into HW13, with guidance from semi-analytic and hydrodynamic galaxy formation models) and study the constraints on \fsig. The full model can be the HOD model or SCAM models \citep[e.g.][]{Guo15b,Guo16}. The model should cover a reasonable range of cosmological parameters and growth rates \citep[e.g.][]{Reid14,DeRose18}. Given the complexity and the demand of parameter exploration, a simulation-based method \citep[e.g.][]{Zheng16} or an emulator \citep[e.g.][]{Wibking17,Zhai18} would be the ideal tool for the investigation. Such an investigation would enable us to quantify the information content of the RSD at different scales and the improvement in the \fsig\ constraints with the small- to intermediate-scale RSD measurements. It would also reveal any systematic bias in \fsig\ constraints when the assembly bias effect is not correctly accounted for and identify the minimum scale for unbiased constraints. Meanwhile, we need to continue the efforts of identifying and quantifying the galaxy assembly bias effect in the observational data \citep[e.g.][]{Lin16} and in the galaxy formation model \citep[e.g.][]{Chaves-Montero16}, which would help further improve the halo model of galaxy clustering. \par 

\section*{Acknowledgements}

We thank Kyle Dawson for useful discussions and the anonymous referee for constructive comments. The work is supported by a seed grant at the University of Utah.
The support and resources from the Center for High Performance Computing at the University of Utah are gratefully 
acknowledged. Z.Z. and H.G. acknowledge the support of the National Science Foundation of China (no. 11828302). 

The MultiDark Database used in this paper and the web application providing online access to it were constructed as part of the activities of the German Astrophysical Virtual Observatory as result of a collaboration between the Leibniz-Institute for Astrophysics Potsdam (AIP) and the Spanish MultiDark Consolider Project CSD2009-00064. The Bolshoi and MultiDark simulations were run on the NASA’s Pleiades supercomputer at the NASA Ames Research Center.

\bibliographystyle{mnras}
\bibliography{rsdAB}

\appendix

\section{Measuring the Volume-Averaged 2PCFs}
\label{sec:A}

When deriving the counterpart of the \fsig\ estimator in configuration space [equation~(\ref{eq:estimator})], 
we encounter two terms not present in the one from power spectrum \citep{Percival09}. They are $\bar{\xi}(r)$ and 
$\bar{\bar{\xi}}(r)$, which are the weighted volume-average of the real-space 2PCF $\xi(r)$ \citep{Hamilton92}, 
with the weight being one and the square of the pair separation, respectively. That is,
\begin{equation}
\bar{\xi}(r) \equiv \frac{3}{r^3}\int^r_0 \xi(r^\prime){r^\prime}^2 {\rm d}r^\prime \label{eq:xiavg}
\end{equation}
and
\begin{equation}
\bar{\bar{\xi}}(r) \equiv \frac{5}{r^5}\int^r_0\xi(r^\prime){r^\prime}^4 {\rm d}r^\prime.\label{eq:xiavgavg}
\end{equation}
To obtain such average 2PCFs, one could measure $\xi(r)$ and compute the above two integrals. However, this would
involve the interpolation between the measurements of $\xi(r)$ at discrete pair separations. In this work, we directly
measure the average 2PCFs by assigning each pair the correct weight. For example, with a periodic box, the 
volume-averaged 2PCF $\bar{\xi}(r)$ can be measured through the \citet{Peebles74} estimator,
\begin{equation}
\bar{\xi}(r) = \frac{N_{\rm d}(<r)}{N_{\rm r}(<r)}-1, \label{eq:xiavgest}
\end{equation}
where $N_{\rm d}(<r)$ [$N_{\rm r}(<r)$] is the number of data-data (random-random) pairs with separations smaller than $r$, 
with each pair weighted by unity. The reason that this gives the expected average can be seen by noting that
theoretically
\begin{equation}
N_{\rm d}(<r) = \int^r_0 \left[1+\xi(r^\prime)\right] \times \frac{1}{2} N \times n\times 4\pi {r^\prime}^2 {\rm d}r^\prime \label{eq:Ngalpair}
\end{equation}
and
\begin{equation}
N_{\rm r}(<r) = \int^r_0 \frac{1}{2}N\times n \times 4\pi {r^\prime}^2 {\rm d}r^\prime = \frac{1}{2} N\times \left(n\times \frac{4}{3}\pi r^3\right), \label{eq:Nrandpair}
\end{equation}
with $N$ and $n$ being the total number and number density of objects in the simulation box. 
Substituting equations~(\ref{eq:Ngalpair}) and (\ref{eq:Nrandpair}) to equation~(\ref{eq:xiavgest}), we obtain 
equation~(\ref{eq:xiavg}). Similarly, the average $\bar{\bar{\xi}}(r)$ can be measured by assigning each pair in 
equation~(\ref{eq:xiavgest}) the square of the pair separation, and the derivation can be seen by multiplying
${r^\prime}^2$ in the integrands of equations~(\ref{eq:Ngalpair}) and (\ref{eq:Nrandpair}). 

\section{Multipoles from Truncated Redshift-Space 2PCFs}
\label{sec:B}

In the Kaiser regime \citep{Kaiser87}, the redshift-space 2PCF $\xi(s,\mu)$ can be decomposed into contributions
from three multipole moments $\xi_l(s)$ ($l=$0, 2, and 4),
\begin{equation}
\xi(s,\mu)=\sum_{l=0,2,4}\xi_l(s) \Pa_l(\mu),
\label{eq:exp}
\end{equation}
where
\begin{equation}
\xi_l(s)= (2l+1)\int^{1}_{0}\xi(s,\mu)\Pa_l(\mu){\rm d}\mu
\end{equation}
and $\Pa_l(\mu)$ is the $l$-th order Legendre polynomial.

The \fsig\ estimator proposed in this work is based on such a decomposition (see \citealt{Percival09}).
At small transverse pair separations, the redshift-space 2PCF is affected by the FOG effect. To reduce such an effect, 
we can limit the calculation to large transverse separations ($\rp>r_{\rm p,cut}$) and compute the modified multipoles
\citep{Reid14,Mohammad16},
\begin{equation}
\hat{\xi}_l(s) = (2l+1)\int_0^{\mu_{\rm max}} \xi(s,\mu)\Pa_l(\mu){\rm d}\mu,
\label{eq:modified}
\end{equation}
where $\mu_{\rm max} = \sqrt{1-({r_{\rm p,cut}/s})^2}$. Substituting the expression of $\xi(s,\mu)$ in equation~(\ref{eq:exp})
into equation~(\ref{eq:modified}), we obtain
\begin{equation}
\boldsymbol{\hat{\xi}}={\sf R}\boldsymbol{\xi},
\end{equation}
where $\boldsymbol{\hat{\xi}}=(\hat{\xi}_0,\hat{\xi}_2,\hat{\xi}_4)^T$, $\boldsymbol{\xi}=(\xi_0, \xi_2, \xi_4)^T$,
and ${\sf R}$ is a 3$\times$3 matrix with elements
\begin{equation}
{\sf R}_{ij}=(2i+1)\int_0^{\mu_{\rm max}}\Pa_i(\mu)\Pa_j(\mu){\rm d}\mu,\:\:\:\:\:    i,j=0,2,4.
\end{equation}
Therefore with the modified multipoles computed from the truncated redshift-space 2PCF, we can obtain the multipoles 
through
\begin{equation}
\boldsymbol{\xi}={\sf R}^{-1}\boldsymbol{\hat{\xi}}.
\end{equation}

\section{Comparison of Colour-Dependent Redshift-Space Clustering Measurements from the Mock and the SDSS Data}
\label{sec:C}

\begin{figure*}
\includegraphics[width=\textwidth]{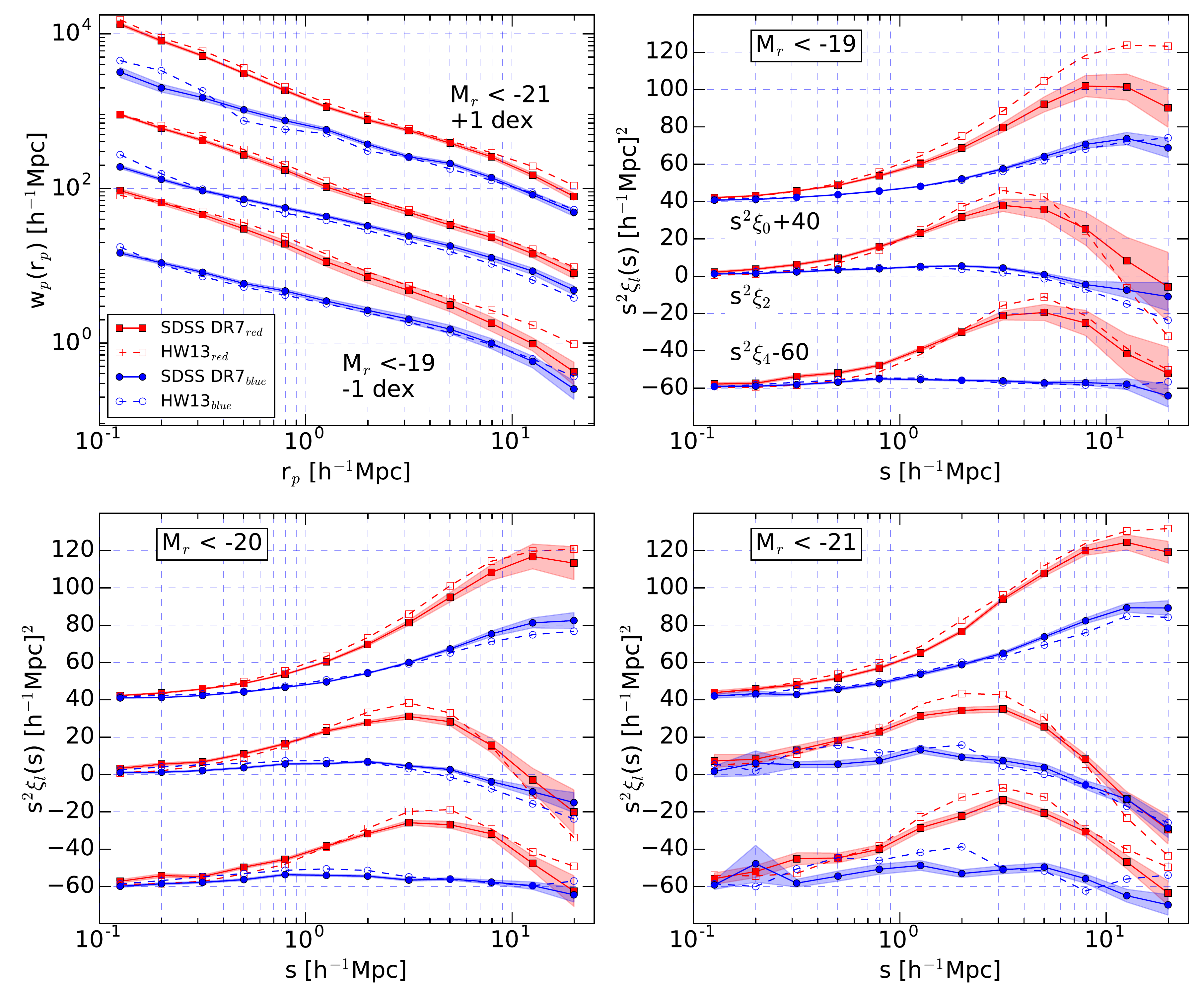}
\caption{
Similar to Fig.~\ref{fig:mlt}, but comparing the colour-dependent 2PCFs and multipoles from the HW13 mock and those
from the SDSS DR7 data.
}
\label{fig:mltdr7blue}
\end{figure*}

In this work, our analyses are focused only on the luminosity-threshold samples of the HW13 galaxy mock. Galaxy
luminosity is assigned using the SHAM method, which places more luminous galaxies into haloes/subhaloes of 
higher \Vpeak\
(with scatter). Therefore, halo assembly bias encoded in \Vpeak\ translates to galaxy assembly bias in the 
luminosity-threshold sample.  Although compared to the SDSS DR7 measurements the mock reproduces the luminosity-dependent 
projected 2PCF reasonably well, we find that there are significant deviations in the redshift-space 2PCF multipoles,
implying that the assembly bias present in the HW13 mock is not realistic.

Assembly bias is also introduced in the HW13 mock through galaxy colour $g-r$, which is assigned through the 
age-matching algorithm. At a fixed luminosity bin, redder colour is assigned to haloes/subhaloes with the higher 
`starvation' redshift (see HW13 for more details). By construction, the distribution of galaxy colour 
at fixed luminosity bin is matched to that of the SDSS galaxies. Here halo assembly bias encoded in `starvation' 
redshift translates to galaxy assembly bias in colour. 

In this appendix, we compare the colour-dependent 2PCFs of
galaxies from the mock and from the SDSS DR7 measurements, with the division between the blue and red galaxy samples 
following the line $g-r=0.21-0.03M_r$. The comparison is shown in Fig.~\ref{fig:mltdr7blue}. We see that for
the projected 2PCFs, there are already clear differences. For red galaxies, on most scales, the projected 2PCFs from 
the HW13 mock are slightly higher than the SDSS DR7 measurements. For blue galaxies, those from the HW13 mock are
lower on scales of a few $\mpchi$ and higher on scales of a few tenth of $\mpchi$. The differences in the 
redshift-space 2PCF multipoles are also clear, with the general trend of an overprediction (underprediction) of
the HW13 mock for red (blue) galaxy multipoles on scales of a few $\mpchi$.  The comparison shows that, like
the luminosity-threshold samples, galaxy assembly bias in terms of galaxy colour in the HW13 mock differs from 
that in the SDSS galaxies (if there is any). 

\section{Best-fitting Model Parameters}
\label{sec:D}

In Tables~\ref{table:hod} and \ref{table:scam}, We list the HOD and SCAM parameters that bestfit the projected correlation function, the redshift-space multipoles, and the number densities for each of the three luminosity-threshold samples in the HW13 mock. 

The HOD model has 7 parameters, the characteristic halo mass scale $M_{\rm min}$ and transition width $\sigma_{\log M}$ for the central galaxy occupation function, the low-mass cutoff $M_0$, amplitude $M_1^\prime$, and slope $\alpha$ for the satellite occupation function, and the central and satellite velocity bias parameters $\alpha_c$ and $\alpha_s$. For each SCAM model, there are 6 parameters --- the counterparts of $M_{\rm min}$ and $\sigma_{\log M}$, denoted as $\mu_{\rm cen}$ and $\sigma_{\rm cen}$; those for satellite galaxies (in sub-haloes), $\mu_{\rm sat}$ and $\sigma_{\rm sat}$; and the central and satellite velocity bias parameters $\alpha_c$ and $\alpha_s$. See more details in section~\ref{sec:modfits}.

\begin{table}
\caption{Best-fitting HOD parameters.} 
\centering 
\begin{tabular}{rrrr} 
\hline \hline                        
Parameters & ${\rm M}_r<-19$ & ${\rm M}_r< -20$ & ${\rm M}_r < -21$ \\ 
\hline 
$\log M_{\rm min}$ & $11.41^{+0.22}_{-0.02}$ & $12.08^{+0.10}_{-0.08}$ & $13.00^{+0.08}_{-0.06}$ \\ 
$\sigma_{\log M}$ & $0.13^{+0.45}_{-0.13}$ & $0.50^{+0.11}_{-0.12}$  & $0.74^{+0.06}_{-0.05}$  \\
$M_0$ & $11.60^{+0.18}_{-0.20}$ & $11.95^{+0.10}_{-0.24}$  &  $11.23^{+0.51}_{-0.98}$ \\
$M_1^\prime$ & $12.85^{+0.02}_{-0.10}$ & $13.35^{+0.69}_{-0.05}$ & $13.98^{+0.05}_{-0.01}$ \\
$\alpha$ & $1.11^{+0.05}_{-0.03}$ & $1.19^{+0.46}_{-0.03}$ & $1.30^{+0.05}_{-0.003}$ \\ 
$\alpha_c$ & $0.44^{+0.01}_{-0.14}$ & $0.30^{+0.03}_{-0.02}$ & $0.22^{+0.09}_{-0.01}$ \\
$\alpha_s$ & $0.67^{+0.03}_{-0.03}$ & $0.76^{+0.01}_{-0.01}$ & $0.80^{+0.01}_{-0.03}$ \\ 
\hline 
\end{tabular}
\label{table:hod} 
\end{table}

\begin{table}
\caption{Best-fitting SCAM parameters.} 
\centering 
\begin{tabular}{rrrr} 
\hline \hline                        
Parameters & ${\rm M}_r< -19$ & ${\rm M}_r <-20$ & ${\rm M}_r< -21$ \\ [0.5ex] 
\hline 
\Vacc\ \, $\mu_{\rm cen}$ & $2.06^{+0.12}_{-0.01}$ & $2.34^{+0.10}_{-0.01}$ & $2.52^{+0.03}_{-0.01}$  \\ 
$\sigma_{\rm cen}$ & $0.07^{+0.19}_{-0.07}$ & $0.23^{+0.04}_{-0.11}$  & $0.21^{+0.33}_{-0.01}$  \\
$\mu_{\rm sat}$ & $2.05^{+0.03}_{-0.01}$ & $2.25^{+0.04}_{-0.01}$  &  $2.46^{+0.04}_{-0.02}$ \\
$\sigma_{\rm sat}$ & $0.01^{+0.07}_{-0.01}$ & $0.01^{+0.07}_{-0.01}$ & $0.13^{+0.05}_{-0.04}$ \\
$\alpha_{c}$ & $0.22^{+0.05}_{-0.16}$ & $0.01^{+0.16}_{-0.01}$ & $0.17^{+0.01}_{-0.04}$ \\ 
$\alpha_{s}$ & $0.88^{+0.01}_{-0.07}$ & $1.00^{+0.0005}_{-0.04}$ & $0.97^{+0.02}_{-0.02}$ \\
\Macc\ \, $\mu_{\rm cen}$ & $12.93^{+0.15}_{-0.74}$ & $12.88^{+0.01}_{-0.16}$ & $12.82^{+0.14}_{-0.01}$  \\ 
$\sigma_{\rm cen}$ & $1.34^{+0.11}_{-0.62}$ & $1.04^{+0.001}_{-0.11}$  & $0.56^{+0.14}_{-0.002}$  \\
$\mu_{\rm sat}$ & $12.12^{+0.08}_{-0.16}$ & $12.31^{+0.02}_{-0.06}$  &  $12.53^{+0.09}_{-0.01}$ \\
$\sigma_{\rm sat}$ & $0.03^{+0.21}_{-0.02}$ & $0.03^{+0.17}_{-0.03}$ & $0.04^{+0.24}_{-0.01}$ \\
$\alpha_{c}$ & $0.18^{+0.18}_{-0.03}$ & $0.15^{+0.01}_{-0.02}$ & $0.13^{+0.04}_{-0.01}$ \\ 
$\alpha_{s}$ & $0.77^{+0.04}_{-0.03}$ & $0.87^{+0.05}_{-0.01}$ & $1.02^{+0.01}_{-0.03}$ \\
\Vpeak\ \, $\mu_{\rm cen}$ & $2.11^{+0.03}_{-0.01}$ & $2.30^{+0.001}_{-0.02}$ & $2.57^{+0.03}_{-0.02}$  \\ 
$\sigma_{\rm cen}$ & $0.07^{+0.06}_{-0.05}$ & $0.15^{+0.002}_{-0.03}$  & $0.21^{+0.04}_{-0.03}$  \\
$\mu_{\rm sat}$ & $2.11^{+0.05}_{-0.0003}$ & $2.29^{+0.03}_{-0.02}$  &  $2.61^{+0.05}_{-0.09}$ \\
$\sigma_{\rm sat}$ & $0.03^{+0.11}_{-0.005}$ & $0.10^{+0.61}_{-0.05}$ & $0.25^{+0.04}_{-0.11}$ \\
$\alpha_{c}$ & $0.11^{+0.10}_{-0.07}$ & $0.01^{+0.10}_{-0.01}$ & $0.02^{+0.16}_{-0.01}$ \\ 
$\alpha_{s}$ & $0.99^{+0.02}_{-0.03}$ & $1.03^{+0.02}_{-0.01}$ & $1.00^{+0.02}_{-0.07}$ \\

\hline 
\end{tabular}
\label{table:scam} 
\end{table}

\bsp    
\label{lastpage}
\end{document}